\documentclass[11pt]{article}

\usepackage[margin=1in]{geometry}
\usepackage{amsmath}
\usepackage{amssymb}
\usepackage{nicefrac}
\usepackage{algorithm}
\usepackage{algorithmic}
\usepackage{xcolor}
\usepackage{tikz}
\usepackage{subcaption}
\usepackage{placeins}
\usepackage{graphicx}
\usepackage[hidelinks]{hyperref}
\usepackage{amsthm}
\usepackage[numbers,square,sort&compress]{natbib}

\newcommand{\rev}[1]{#1}

\newenvironment{revision}{}{}

\newtheorem{lemma}{Lemma}
\newtheorem{definition}{Definition}
\newtheorem{remark}{Remark}
\newtheorem{corollary}{Corollary}
\newtheorem{observation}{Observation}

\newcommand{\voters}{\mathcal{V}}
\newcommand{\alternatives}{\mathcal{A}}
\newcommand{\voter}{v}
\newcommand{\alternative}{a}
\newcommand{\ranklist}{\sigma}
\newcommand{\votingrule}{\mathcal{M}}
\newcommand{\util}{u}
\newcommand{\cost}{c}
\newcommand{\plu}[2]{|\voters^#1(#2)|}

\newcommand{\SW}{\mathsf{SW}}
\newcommand{\SC}{\mathsf{SC}}

\newcommand{\distortion}{\mathsf{D}}

\title{Distortion in Online Voting}

\author{
	MohammadHossein Ghasemi\thanks{Chinese University of Hong Kong, Shenzhen.}
	\and
	Masoud Seddighin\thanks{Tehran Institute for Advanced Studies.}
	\and
	Saeed Seddighin\thanks{Toyota Technological Institute at Chicago.}
}

\date{}

\begin{document}
	
	\maketitle
	
	\begin{abstract}
		We initiate the study of distortion in online voting, a framework that bridges computational social choice and online algorithms. In contrast to the classical offline setting where the full election instance is known in advance, we consider a dynamic environment where a fixed set of voters is present from the beginning, but alternatives arrive sequentially over time. Upon each arrival, a mechanism must immediately and irrevocably decide whether to retain the currently selected candidate or switch to the newly arriving one, operating solely on the partial ordinal rankings induced by the underlying cardinal preferences.

We provide a collection of upper and lower bounds for both deterministic and randomized mechanisms under the two standard cardinal preference frameworks:
\begin{itemize}
\item \textbf{Utilitarian Framework:} For deterministic mechanisms under an adversarial arrival order, we show a separation from the offline setting by establishing a tight lower bound of $\Omega(mn)$, which we match with an online algorithm achieving a distortion of $\frac{2}{3}m(n+\frac{1}{4})$. For randomized rules, we introduce a regularity property showing that any \emph{regular} offline rule can be converted into an online rule with equal expected distortion. We prove that the Harmonic voting rule satisfies this property, resulting in an online randomized mechanism with $O(\sqrt{m\log m})$ distortion.
\item \textbf{Metric Framework:} For deterministic mechanisms, we show that maintaining an alternative within a $\varphi$-weighted uncovered set guarantees a distortion of $2+\sqrt{5}$, and prove that the state-of-the-art offline rules achieving distortion $3$ cannot be directly adapted online. We establish a deterministic lower bound of $\sqrt{13}$, even for three alternatives, and give a matching upper bound of $\sqrt{13}$ for three alternatives and any number of voters. For randomized mechanisms, we use regularity to show that Random Dictatorship guarantees an online distortion of $3-\frac{2}{n}$.
\end{itemize}

	\end{abstract}
	
	
	\section{Introduction}

In computational social choice, an election selects a single candidate from a set of alternatives by aggregating the preferences of voters \cite{sen1986social}. Although these preferences are often rich and complex, voting rules typically rely on simple elicitation formats. The most common approach asks each voter to submit a linear order over the alternatives, and the winner is determined solely from these rankings.

Following classical decision-theoretic models \cite{neumann1944theory}, each agent assigns a cardinal value to every alternative, representing either its cost or its utility. If these values represent costs, the socially optimal alternative is the one that minimizes the total cost across all agents. If they represent utilities, the socially optimal alternative is the one that maximizes the total utility.

Since ordinal rankings retain only the relative order of alternatives and discard cardinal information, voting rules based solely on rankings may select outcomes that are far from optimal under the underlying cardinal preferences. \rev{This inefficiency is captured by the notion of distortion. For costs, distortion compares the cost of the selected alternative with the minimum possible cost; for utilities, it compares the maximum possible welfare with the welfare of the selected alternative. In both cases, the ratio is evaluated in the worst case over cardinal preferences consistent with the observed rankings.}

The notion of distortion was introduced by Procaccia and Rosenschein \cite{procaccia2006distortion} and has since become a central measure of efficiency loss in social choice \cite{anshelevich2021distortion}. This notion has been studied under a variety of models and assumptions, and the studies have emerged within two main frameworks: the utilitarian framework \cite{procaccia2006distortion} and the metric framework \cite{anshelevich2015approximating}.
 In the utilitarian framework, each agent is endowed with a utility function over alternatives, typically normalized so that utilities sum to one. In the metric framework, agents and alternatives are embedded in a metric space, and each agent’s cost for an alternative is given by the distance between them. In both settings, the voting rule observes only the induced ordinal rankings and does not have access to the underlying cardinal values.

\rev{The offline setting admits strong distortion guarantees in both frameworks.} In the utilitarian setting, no deterministic voting rule can achieve distortion better than $\Omega(m^2)$, where $m$ is the number of alternatives, and this bound is tight via simple rules such as Plurality \cite{caragiannis2017subset,procaccia2006distortion}. In the metric setting, no deterministic voting rule can achieve distortion better than $3$ \cite{anshelevich2015approximating}. The first constant-distortion result showed that the Copeland rule achieves distortion $5$ \cite{anshelevich2015approximating}. A sequence of more sophisticated voting rules subsequently reduced this guarantee to $4.236$ \cite{Munagala2019Improved} and finally to the optimal bound of $3$ \cite{gkatzelis2020resolving,kizilkaya2022pluralityveto}. \rev{For randomized rules, the optimal distortion is $\Theta(\sqrt{m})$ in the utilitarian framework \cite{boutilier2015optimal,ebadian2024optimized}. In the metric framework, a lower bound of $2.1126$ is known \cite{charikar2022metric}, while successive works improved the upper bound from $2.753$ \cite{charikar2024breaking} to $2.5$ \cite{frank2026improved,ye2026half}, $2.1441$ \cite{shah2026improving}, and $2.13713$ \cite{cai2026stable}.}

\rev{In this work, we study distortion in \textbf{the online voting model}.}

\subsection{Online Voting}
Online algorithms are among the most extensively studied areas of algorithm design. In this setting, the input is revealed over time rather than given in advance. As new information arrives, the algorithm must make decisions immediately and irrevocably, without knowledge of future inputs. This framework naturally models many real-world scenarios where waiting to observe the full input is impossible or undesirable.

In contrast, most work in voting theory assumes an offline setting, where the full election—both the set of candidates and the preferences of all voters—is known in advance. This assumption is appropriate for traditional elections, where the set of candidates is fixed before voting begins. However, many modern applications of voting are inherently online.
For example, in recommendation systems, new items arrive continuously, and the system must repeatedly decide which item to promote based on current user feedback, without knowing future items. Similarly, in hiring or admission processes, candidates may apply over time, and decisions or rankings may need to be updated as new applications arrive rather than after all applications are received. In crowdsourcing platforms, proposals or tasks can appear sequentially, and a selection must be made without access to all future submissions.

 In an online voting problem, two components may be revealed gradually: the set of voters and the set of alternatives. \rev{We focus on the setting where all voters are present from the beginning and only the candidates arrive over time.}
 
In this paper, we consider a setting in which alternatives arrive online, and at each arrival the voting system must decide whether to keep the currently selected candidate or replace it with the newly arriving one. This decision is irrevocable: once an alternative is rejected, it cannot be reconsidered later. At each time step, each voter reports a ranking over the set of currently revealed alternatives, and the decision is made based only on these partial rankings. \rev{This replacement mechanism follows the established online model with \emph{preemption}, also called \emph{free disposal}, studied in online optimization \cite{buchbinder2015preemption,chan2017freedisposal}; here the maintained solution consists of a single candidate.}

As an example, suppose a committee of voters is tasked with hiring a single candidate from a pool of applicants who arrive one by one. After each interview, the committee must either \rev{reject the arriving candidate permanently and keep the previously selected candidate}, or replace the current winner with the arriving candidate and permanently discard the previous one. The goal is to ultimately select a high-quality candidate despite having access only to partial, sequential information.

These scenarios naturally raise the following question: how much additional distortion is incurred when a voting rule must operate in an online environment?

\begin{remark}
Our online voting model is closely related to the classical secretary problem. One can view the alternatives as job candidates arriving one by one, while the hiring committee (the voters) remains fixed and reveals its preferences over the candidates seen so far. As each new candidate arrives, the committee decides whether to keep its current favorite or replace it with the newcomer.
\rev{In the classical secretary problem, the algorithm may accept at most one candidate and cannot later replace an accepted candidate. In our model, the current winner may be replaced, but a rejected alternative remains unavailable. Thus rejection is irreversible in both settings; the difference is the ability to replace the selected alternative. We allow both the arrival order and the underlying cardinal values to be adversarial. An interesting direction for future work is to study online ordinal algorithms under random arrival orders or stochastic assumptions on the cardinal values.
}\end{remark}

\begin{revision}
\subsection{Our Results}
We provide upper and lower bounds for deterministic and randomized online voting rules under both cardinal preference frameworks. Throughout, $n$ denotes the number of voters and $m$ the number of alternatives.

In the utilitarian framework, we show that every deterministic online rule has distortion at least \rev{$\frac23m(n-\frac12)$} for \rev{$n\ge2$ a power of two and $m\ge\log_2 n+2$}, and give an online algorithm with distortion at most $\frac23m(n+\frac14)$ for $n\ge2$. Compared with the offline $O(m^2)$ guarantee, these bounds give an asymptotic separation as $n/m$ grows within the described range. \rev{For $n=2^k$ with $k\ge1$ and any $m\ge2$, we also prove a lower bound of $\frac{mn}{2\max\{1,2^{\log_2 n-m+1}\}}+1$.}

For randomized rules, we identify a property, called regularity, that allows us to convert any order-independent offline rule satisfying it into an online rule with the same outcome distribution at every prefix and hence the same distortion. Regularity requires that adding an alternative cannot increase the selection probability of an existing alternative. Applying this conversion to the Harmonic rule gives online utilitarian distortion $O(\sqrt{m\log m})$.

In the metric framework, we prove a deterministic lower bound of $\sqrt{13}$ over electorate sizes, even with three alternatives, separating online voting from the optimal offline guarantee of $3$. We give a matching upper bound for three alternatives that holds for every number of voters, establishing $\sqrt{13}$ as the optimal uniform bound. For an arbitrary number of alternatives, we give a deterministic online rule with distortion at most $2+\sqrt5$. We also show that matching-based rules that achieve distortion $3$ in the offline setting cannot be directly implemented online.
Finally, we show that Random Dictatorship satisfies regularity and yields online metric distortion at most $3-2/n$. \end{revision}

	\section{Further Related Work}
Most of the main related results have already been discussed in the introduction. Here, we briefly mention additional related work. For a broader overview of distortion, we refer to the survey by Anshelevich et al.~\cite{anshelevich2021distortion}.

For randomized rules, several early results were obtained in the metric setting. Anshelevich and Postl~\cite{anshelevich2017randomized} showed that Random Dictatorship achieves distortion $3 - 2/n$ and proved a general lower bound of $2$ for all randomized rules. Kempe~\cite{kempe2020communication} improved the upper bound to $3 - 2/m$. Charikar et al.~\cite{charikar2024breaking} later obtained an upper bound of $2.753$, which was subsequently improved to $2.5$ \cite{frank2026improved,ye2026half}, $2.1441$ \cite{shah2026improving}, and $2.13713$ \cite{cai2026stable}.

Beyond single-winner elections, distortion has also been studied in committee selection, where the goal is to choose a set of $k>1$ winners. This line of work was initiated by Caragiannis et al.~\cite{caragiannis2017subset}. Subsequent research has developed both utilitarian and metric variants under different objective functions, see, e.g., \cite{goel2018relating,caragiannis2022metric,kizilkaya2022pluralityveto,elkind2017properties,faliszewski2017multiwinner,chen2020favorite,cheng2020group,kalayci2024proportional}.

Distortion has also been studied under various additional modeling assumptions. Cheng et al.~\cite{cheng2017people,cheng2018distortion} considered the case where candidates are drawn from the population of voters. Ghodsi et al.~\cite{ghodsi2019distortion} studied settings in which voters are allowed to abstain.

Finally, this work is related to social choice with varying candidate sets. Laslier~\cite{laslier2000aggregation} studied cloning consistency, which requires that replication of alternatives does not affect the outcome. In contrast, we consider a sequential model in which alternatives arrive over time and rejected candidates cannot be reconsidered.

	\section{Preliminaries}
\label{sec:preliminaries}

In an online voting problem, we are given a set
$\mathcal{V} = \{\voter_1, \voter_2, \ldots, \voter_n\}$
of $n$ voters and a set
$\mathcal{A} = \{\alternative_1, \alternative_2, \ldots, \alternative_m\}$
of $m$ alternatives. Each voter \( \voter_j \in \mathcal{V} \) has a strict linear order \( \ranklist_j \) over the alternatives, representing her preferences from most to least preferred. We say that voter $\voter_j$ prefers alternative $\alternative_i$ to alternative $\alternative_k$, denoted by $\alternative_i \succ_j \alternative_k$, if alternative $\alternative_i$ appears before alternative $\alternative_k$ in $\voter_j$'s ranklist.

\paragraph{Online setting.} In the online setting, alternatives arrive sequentially. We assume without loss of generality that alternatives arrive in the order $\alternative_{1},\alternative_{2}, \ldots, \alternative_m$. At each step \(t\), a new alternative \(\alternative_t\) arrives. For each voter \( \voter_j \), we observe \( \ranklist_j^t \), which is her ranking restricted to the first \(t\) alternatives.

An online voting mechanism begins by selecting \(\alternative_1\) as the initial winner. At each subsequent step \(t \in \{2,3,\ldots,m\}\), the mechanism observes the partial ranklists for all voters and decides whether to retain the current winner or replace it with the newly arrived alternative \(\alternative_t\). This decision is irrevocable: once the winner is replaced, previously discarded alternatives cannot be reconsidered. The winner after step \(m\) is the final outcome of the election.

\begin{definition}\label{def:votersets}
	At step \(t\), let \( \voters^{t}(\alternative_i) \) denote the set of voters who rank alternative \( \alternative_i \) in the first position of their partial ranking. 
		Moreover, denote by \( \voters_{i \succ k} \) the set of voters who prefer \( \alternative_i \) to \( \alternative_k \). More generally, for a subset $X$ of alternatives, let $\voters_{i \succ X}$ denote the set of voters who strictly prefer alternative $\alternative_i$ to every alternative in set $X$.
\end{definition}

Notice that in Definition \ref{def:votersets}, the first set depends on the current step \(t\), whereas the second set is independent of \(t\) and is determined solely by the pairwise comparison between \(\alternative_i\) and \(\alternative_k\).

\paragraph{Utilitarian and Metric frameworks.} Our assumption is that the ranklists are induced by underlying numerical values, interpreted either as utilities or costs. Two standard frameworks capture this assumption: the utilitarian framework and the metric framework.
In the utilitarian framework, each voter \( \voter_j \) assigns a nonnegative utility \( \util_{i,j} \) to every alternative \( \alternative_i \), representing the value of selecting \( \alternative_i \) as the winner. The utilities are normalized so that, for every voter \( \voter_j \),
$
\sum_{i=1}^{m} \util_{i,j} = 1.
$

In the metric framework, voters and alternatives are embedded in a metric space \( (\mathcal{X}, d) \), where \( d : \mathcal{X} \times \mathcal{X} \to \mathbb{R}_{\geq 0} \) is a distance function satisfying, for all \( x, y, z \in \mathcal{X} \),
\[
\begin{aligned}
d(x,y)=0&\iff x=y, \qquad d(x,y)=d(y,x),\\
d(x,y)&\le d(x,z)+d(z,y).
\end{aligned}
\]
The cost incurred by voter $\voter_j$ for alternative $\alternative_i$ is defined as $\cost_{i,j} = d(\alternative_i,\voter_j)$. 

Each voter determines her ranklist by sorting the alternatives according to the induced preference values. In the metric framework, alternatives are ordered in increasing order of cost, i.e.,
$
\cost_{i,j} < \cost_{k,j}$ implies $\alternative_i \succ_j \alternative_k.
$
In the utilitarian framework, alternatives are ordered in decreasing order of utility, i.e.,
$
\util_{i,j} > \util_{k,j}$ implies $\alternative_i \succ_j \alternative_k.
$
Ties in both frameworks are broken arbitrarily but consistently across prefixes. 
Note that these costs and utilities are not observable to the voting mechanism; at each step, the mechanism makes its decision solely based on the observed ranklists.

In the utilitarian framework, the social welfare of an alternative $\alternative_i$ is defined as the sum of its utilities across all voters: $\SW(\alternative_i) = \sum_{j=1}^{n} \util_{i,j}$. In the metric framework, the total cost of an alternative $\alternative_i$ is defined analogously as the sum of its distances to all voters: $\SC(\alternative_i) = \sum_{j=1}^{n} \cost_{i,j}$.

\paragraph{Distortion.}

The distortion of a voting mechanism \(\mathcal{M}\) on an instance \(\mathcal{E}\) measures how inefficient the chosen alternative is compared to the optimal one.

In the utilitarian framework, consider an online voting problem \(\mathcal{E} = (\mathcal{V}, \mathcal{A}, \mathbf{U})\), where \(\mathbf{U}\) is an \(m \times n\) utility matrix. The distortion of \(\mathcal{M}\) on \(\mathcal{E}\) is defined as
\[
\distortion(\mathcal{M}, \mathcal{E}) = \frac{\SW(\alternative^*)}{\SW(\alternative_w)},
\]
where
$
\alternative^* \in \arg\max_{\alternative_i \in \mathcal{A}} \SW(\alternative_i)
$
is the socially optimal alternative, and \(\alternative_w\) is the alternative selected by \(\mathcal{M}\).

For the metric framework, the definition is analogous. Given an online voting problem \(\mathcal{E} = (\mathcal{V}, \mathcal{A}, \mathbb{C})\), where \(\mathbb{C}\) is an \(m \times n\) cost matrix, the distortion of \(\mathcal{M}\) on \(\mathcal{E}\) is defined as
\[
\distortion(\mathcal{M}, \mathcal{E}) = \frac{\SC(\alternative_w)}{\SC(\alternative^*)},
\]
where
$
\alternative^* \in \arg\min_{\alternative_i \in \mathcal{A}} \SC(\alternative_i)
$
is the socially optimal alternative, and \(\alternative_w\) is the alternative selected by \(\mathcal{M}\).

For a randomized mechanism, let $W$ denote its final winner. The corresponding definitions are
\[
\distortion_{\mathrm{util}}(\mathcal{M},\mathcal{E})
=\frac{\SW(\alternative^*)}{\mathbb{E}[\SW(W)]},
\qquad
\distortion_{\mathrm{metric}}(\mathcal{M},\mathcal{E})
=\frac{\mathbb{E}[\SC(W)]}{\SC(\alternative^*)}.
\]
The expectation is over the mechanism's internal randomness; the full preference profile and arrival order are fixed independently of these random choices. The unit-sum normalization applies to the full utility profile, rather than separately to each prefix. A positive numerator with a zero denominator gives infinite distortion; we assign the ratio $1$ when both are zero.

In both frameworks, the distortion of a voting rule $\votingrule$ is then defined as the worst-case distortion over all possible elections:
\[
\distortion(\votingrule) = \sup_{\mathcal{E} \in \Omega} \distortion(\votingrule, \mathcal{E}),
\]
where $\Omega$ denotes all instances and arrival orders with $n$ voters and $m$ alternatives in the respective framework.

	\section{Utilitarian Framework}
\label{sec:util}
We begin by establishing a collection of possibility and impossibility results for the utilitarian framework. First, we prove a lower bound on the distortion of any deterministic online voting rule, showing that it is at least
$
\nicefrac{2}{3}mn-\rev{\nicefrac{1}{3}}m.
$
The proof is based on a construction in which, for any online algorithm, the selected alternative is \rev{ranked last among the alternatives revealed when it arrives by $n-1$ voters}, while another alternative is ranked first by at least half of the voters. We note that this construction requires \rev{$m\ge\log_2 n+2$ for $n\ge2$ a power of two} and therefore does not apply to all parameter regimes.


\begin{lemma}[Utilitarian, deterministic, adversarial order] \label{lem:util-det-adv-2}
	\rev{Let $n\ge2$ be a power of two and $m\ge\log_2 n+2$.}
	The distortion of any deterministic voting rule for the online voting problem under the utilitarian assumption and adversarial arrival order of alternatives is at least
	$
	\nicefrac{2}{3}m\left(n-\nicefrac{1}{2}\right).
	$
\end{lemma}
\begin{proof}
	
	Let \rev{$n\ge2$ be a power of two and assume $m\ge\log_2 n+2$}. At step \(2\), suppose that \( \alternative_1 \) is ranked first by exactly \( \nicefrac{n}{2} \) voters, while \( \alternative_2 \) is ranked first by the remaining \( \nicefrac{n}{2} \) voters. The algorithm must choose one of these two alternatives. \rev{We describe the case in which it selects $\alternative_1$. If it selects $\alternative_2$, interchange their roles and the two voter groups in the construction below; the argument allows either decision at every subsequent step.}

	We describe the adversarial strategy for the arrival of subsequent alternatives. Let $\alternative_w$ be the alternative selected by the algorithm at the end of step $t-1$. At step $t$, the adversary introduces alternative $\alternative_t$ such that it is ranked above $\alternative_w$ by $\rev{\frac{\plu{{t-1}}{\alternative_w}}{2}}$ voters from the \(\voters^{t-1}(\alternative_w)\), and is ranked last by all other voters. 
	
	The algorithm now faces the following choice:
	
	\begin{itemize}
		
		\item If it switches to \( \alternative_t \), then \(\rev{|\voters^t(\alternative_t)|} =\rev{\frac{\plu{{t-1}}{\alternative_w}}{2}}\) \rev{voters rank} the selected alternative first, while all other voters rank \( \alternative_t \) last.
		
		\item If it keeps \( \alternative_w \), then its number of first-place rankings decreases to \(\rev{|\voters^t(\alternative_w)|} = \rev{\frac{\plu{{t-1}}{\alternative_w}}{2}}\).
		
	\end{itemize}
	
	To illustrate, \rev{for $n\ge4$,} consider step $3$ when $\alternative_3$ arrives. Suppose that under the current partial rankings, $\alternative_3$ is ranked above $\alternative_1$ by voters $\voter_1, \dots, \voter_{\rev{n/4}}$, and is ranked last by all others. The induced rankings at step $3$ would be:
	
	\[
	\begin{array}{c|c|c|c}
		
		& \voter_1\cdots\voter_{\rev{n/4}} & \voter_{\rev{n/4} + 1}\cdots\voter_{n/2} & \voter_{n/2+1}\cdots\ \voter_n \\ \hline
		
		\text{1st} & \alternative_3 & \alternative_1 & \alternative_2 \\
		
		\text{2nd} & \alternative_1 & \alternative_2 & \alternative_1 \\
		
		\text{3rd} & \alternative_2 & \alternative_3 & \alternative_3
		
	\end{array}
	\]
	
	Since \rev{$m\ge\log_2 n+2$}, we continue this process until the selected alternative $\alternative_w$ is ranked first by exactly one voter; call that voter $\voter_j$. Assume this first occurs at step $t_1-1$. We then introduce a new alternative $\alternative_{t_1}$, place it at the top of $\voter_j$'s ranking (above $\alternative_w$), and place it last for all other voters. All subsequently arriving alternatives are placed last by every voter. \rev{There are $\log_2 n-1$ halving steps, in addition to the initial two alternatives and the final arrival at $t_1$.}
	By step $t_1$, the algorithm must have either switched to $\alternative_{t_1}$ or be left with an alternative that no voter ranks first. If the algorithm does not switch to an alternative with at least one first-place vote, we can define a utility profile where each voter assigns utility $1$ to their top alternative and $0$ to others. Under this assignment, the selected alternative has total welfare \(0\), while another alternative has strictly positive welfare, implying that the distortion becomes unbounded.
	
	Therefore, to maintain a bounded distortion, the algorithm must \rev{select alternative $\alternative_{t_1}$ at step $t_1$ and retain it}. In this construction, $\alternative_{t_1}$ is ranked first by exactly one voter ($\voter_j$), and is ranked last by all other voters \rev{at its arrival}. Conversely, $\alternative_2$ is ranked first by half of the voters ($n/2$). We construct a consistent utility profile as follows:
	
	\begin{itemize}
		\item For voter $\voter_j$, assign utility $1/m$ to every alternative.
		\item For voters $\voter_{n/2+1}, \dots, \voter_n$, assign utility $1$ to top\rev{-ranked} alternative ($\alternative_2$) and $0$ to all others.
		\item For each remaining voter, let $k$ be the rank of $\alternative_2$ in their preference order, and assign utility $1/k$ to each of their top $k$ alternatives\rev{, and $0$ to all others}.
	\end{itemize}
	\begin{revision}
		Consider the last switch before $t_1$, when the current winner has $s$ supporters who rank $\alternative_2$ at position $r$. Since $n$ is a power of two, both halves have size $s/2$. After switching, $s/2$ voters keep $\alternative_2$ at rank $r$, while the other $s/2-1$ voters, excluding $\voter_j$, eventually rank it at $r+2$. Their total contribution is
		\[
		\frac{s/2}{r}+\frac{s/2-1}{r+2}\ge\frac{s-1}{r+1},
		\]
		which is the contribution obtained by retaining the incumbent until $t_1$. The contribution $1/m$ of $\voter_j$ is the same in both cases. Removing the last switch from a decision sequence therefore cannot increase $\SW(\alternative_2)$. Repeating this comparison reduces to the sequence with no switches before $t_1$, where the $n/2-1$ voters other than $\voter_j$ each contribute $1/3$.
	\end{revision}
	
	Under \rev{the above} utility profile, the total social welfare of $\alternative_{t_1}$ is at most $\frac{1}{m}$, since only $\voter_{j}$ contributes $\frac{1}{m}$ to every alternative and all other voters contribute $0$ to $\alternative_{t_1}$.
	On the other hand, alternative $\alternative_2$ receives utility $1$ from each of the $\frac{n}{2}$ voters $\voter_{n/2+1}, \ldots, \voter_n$, and at least utility $\frac{1}{3}(\frac{n}{2}-1) + \frac{1}{m}$ from the remaining voters. Thus, the total social welfare of $\alternative_2$ is
	
	\begin{align*}			
		\SW(\alternative_2) &\ge \frac{n}{2} \cdot 1 + \left(\frac{n}{2} - 1\right) \cdot \frac{1}{3} + \frac{1}{m}\\
		&\ge \frac{2n-1}{3}.		
	\end{align*}

	Therefore, the distortion satisfies
	
	\[
	\distortion \ge \frac{\SW(\alternative_2)}{\rev{\SW(\alternative_{t_1})}} \ge \frac{\frac{2n-1}{3}}{\frac{1}{m}} = \frac{2}{3}m\left(n - \frac{1}{2}\right).
	\]
	
	\rev{For a fixed deterministic rule, the construction determines a fixed instance and arrival order.}
\end{proof}

Notice that the bound established in Lemma~\ref{lem:util-det-adv-2} only applies when \rev{$n\ge2$ is a power of two and $m\ge\log_2 n+2$}. Thus, instances with a smaller number of alternatives could potentially admit better distortion guarantees. To address this, we prove another lower bound in Lemma~\ref{lem:util-det-adv} that works for \rev{all $m\ge2$ when $n$ is a power of two}.

\begin{lemma}[Utilitarian, deterministic, adversarial order] \label{lem:util-det-adv}
	\rev{Let $n=2^k$ for an integer $k\ge1$ and let $m\ge2$.}
	The distortion of any deterministic voting mechanism for the online voting problem with utilitarian assumption and adversarial order of alternatives is lower bounded by $$\frac{mn}{2\max(1, 2^{\rev{\log_2(n)-m+1}})}+1.$$
\end{lemma}
\begin{proof}
	Assume \( n = 2^k \). We construct the preference profile inductively. For each voter \( \voter_j \), consider \( m \) positions labeled from \(1\) (top rank) to \(m\) (lowest rank), representing their ranking. A position is called \emph{blocked} if an alternative has already been assigned to it.
	
	At step \(2\), we compare \( \alternative_1 \) and \( \alternative_2 \). Voters \( \voter_1, \ldots, \voter_{n/2} \) prefer \( \alternative_1 \), while the remaining voters prefer \( \alternative_2 \). \rev{We describe the case in which the rule selects $\alternative_1$; if it selects $\alternative_2$, interchange the two alternatives and voter groups in the remainder of the construction.}
	We then fix the position of \( \alternative_2 \) in each voter’s ranking so that these preferences are respected. For voters \( \voter_1, \ldots, \voter_{n/2} \), we place \( \alternative_2 \) in position \(m\) (the lowest rank). For voters \( \voter_{n/2+1}, \ldots, \voter_n \), we place \( \alternative_2 \) in position \(1\) (the top rank). 
	This assignment is consistent with the revealed pairwise preferences and does not restrict how the remaining positions can be filled in later steps.
	
	\[
	\begin{array}{c|c|c}
		& \voter_1\cdots\voter_{n/2} & \voter_{n/2+1}\cdots\voter_n \\ \hline
		
		\text{1st} & \cdot & \alternative_2 \\
		
		\text{2nd} & \cdot & \cdot \\
		
		\vdots & \vdots & \vdots \\
		
		\text{m-th} & \alternative_2 & \cdot
	\end{array}
	\]
	Now consider step \(3\)\rev{ for $n\ge4$ and $m\ge4$, to illustrate the general construction}. The voters are partitioned into four equal groups:
	\[
	\voter_1,\ldots,\voter_{n/4}, \quad
	\voter_{n/4+1},\ldots,\voter_{n/2}, \quad
	\voter_{n/2+1},\ldots,\voter_{3n/4}, \quad
	\voter_{3n/4+1},\ldots,\voter_n.
	\]
	Voters in the first and third groups prefer \( \alternative_1 \), while voters in the second and fourth groups prefer \( \alternative_3 \).
	
	Since the position of \( \alternative_2 \) was fixed in the previous step (at the top for the second half of voters and at the bottom for the first half), it does not affect the comparison between \( \alternative_1 \) and \( \alternative_3 \).
	\rev{For this illustration, suppose that the voting rule selects $\alternative_3$. If it retains $\alternative_1$, we instead assign $\alternative_3$ to the extremal unblocked positions according to the same rule described below.}
	
	We now fix the position of \( \alternative_1 \). For voters \( \voter_1, \ldots, \voter_{n/4} \) and \( \voter_{n/2+1}, \ldots, \voter_{3n/4} \), 
	we place \( \alternative_1 \) in the smallest unblocked position. 
	For voters \( \voter_{n/4+1}, \ldots, \voter_{n/2} \) and \( \voter_{3n/4+1}, \ldots, \voter_n \), 
	we place \( \alternative_1 \) in the largest unblocked position.
	The resulting partial profile is:
	\[
	\begin{array}{c|c|c|c|c}
		& \voter_1\ldots\voter_{n/4} & \voter_{n/4+1}\ldots\voter_{n/2} & \voter_{n/2+1}\ldots\voter_{3n/4} & \voter_{3n/4+1}\ldots\voter_n \\ \hline
		
		\text{1st}   & \alternative_1 & \cdot & \alternative_2 & \alternative_2 \\
		
		\text{2nd}   & \cdot          & \cdot & \alternative_1 & \cdot \\
		
		\vdots       & \vdots         & \vdots & \vdots        & \vdots \\
		
		\text{m-1}   & \cdot          & \alternative_1 & \cdot & \cdot \\
		
		\text{m-th}  & \alternative_2 & \alternative_2 & \cdot & \alternative_1
	\end{array}
	\]
	
	The above construction naturally extends to all subsequent steps. For a general step \rev{$3\le t\le h:=\min\{m,k+1\}$}, we partition the voters according to the first \rev{$t-2$} bits of their binary indices. More precisely, write each voter index \(j \in [n]\) in binary as
	\[
	\rev{j-1} = (b_1(j), b_2(j), \ldots, b_k(j)),
	\]
	where \(b_1(j)\) is the most significant bit. For each prefix \((b_1,\ldots,b_{\rev{t-2}}) \in \{0,1\}^{\rev{t-2}}\), let
	\[
	B_{b_1,\ldots,b_{\rev{t-2}}} = \{ j \in [n] : b_s(j) = b_s \text{ for all } s < \rev{t-1} \}
	\]
	denote the corresponding block of voters. Each such block is further divided into two equal subblocks according to the \rev{bit $t-1$}, namely \(B_{b_1,\ldots,b_{\rev{t-2}},0}\) and \(B_{b_1,\ldots,b_{\rev{t-2}},1}\).
	
	At step \(t\), a new alternative \(\alternative_t\) is compared with the current winner \(\alternative_w\). Within each pair of subblocks \(B_{b_1,\ldots,b_{\rev{t-2}},0}\) and \(B_{b_1,\ldots,b_{\rev{t-2}},1}\), voters in one subblock prefer \(\alternative_t\) over \(\alternative_w\), while voters in the other subblock prefer \(\alternative_w\) over \(\alternative_t\). This ensures that the comparison is balanced within each pair.
	
	Since all previously assigned alternatives occupy only extremal positions and are placed identically within corresponding subblocks, \rev{both decisions leave a consistent continuation.} We then place the losing alternative \rev{$\alternative_\ell$} in extremal unblocked positions: in subblocks where voters prefer \rev{$\alternative_\ell$ to the retained alternative}, we assign it to the smallest available position, and in the other subblocks to the largest available position.
	
	Note that after step \(t\), the number of voters whose first position is still available decreases exponentially with \(t\). In particular, each step fixes the first position for exactly half of the remaining active blocks. Hence, the number of voters for which the first position is still not blocked after step \(t\), \rev{for $2\le t\le h$,} is
	$
	\rev{2^{\log_2 n - t+1}}.
	$
	\begin{revision}
		\paragraph{Completion of the construction.}
		At step $h$, place the current winner in the highest unblocked position of each voter and fill the other unblocked positions with the remaining alternatives in arrival order. There are $n/2$ unblocked first positions after step $2$, and this number halves at every step through $h$; thus $s=n/2^{h-1}=\max\{1,2^{\log_2 n-m+1}\}$ voters rank this winner first. Every later alternative is below it for every voter. A rule with finite distortion must retain it: after switching to a later alternative, all available outcomes have no first-place supporters, and assigning utility $1$ only to each voter's top alternative gives infinite distortion. This also covers $m=2$ and $n=2$, when no further halving step is needed.
	\end{revision}
	Therefore, after \(m\) steps, the current winner is ranked first by only
	$
	\max\left\{1,\, 2^{\rev{\log_2 n-m+1}}\right\}
	$
	voters.
	
	We complete the proof by defining a utility profile consistent with the constructed ranklists.
	For every voter who ranks the final winner \( \alternative_w \) in the first position, we assign utility \(1/m\) to all alternatives. For every other voter, we assign utility \(1\) to their top-ranked alternative and utility \(0\) to all remaining alternatives.
	Consider \( \alternative_2 \), which is ranked first by exactly \(n/2\) voters. The welfare of \( \alternative_2 \) is
	\[
	\SW(\alternative_2)
	=
	\frac{n}{2} + \frac{1}{m}\max\{1,\, 2^{\rev{\log_2 n-m+1}}\}.
	\]
	
	Now consider the final winner \( \alternative_w \). By construction, \( \alternative_w \) is ranked first by exactly
	$
	\max\{1,\, 2^{\rev{\log_2 n-m+1}}\}
	$
	voters, each contributing \(1/m\). Hence,
	\[
	\SW(\alternative_w)
	=
	\frac{1}{m}\max\{1,\, 2^{\rev{\log_2 n-m+1}}\}.
	\]
	
	Therefore, the distortion is \rev{at least}
	\[
	\frac{\SW(\alternative_2)}{\SW(\alternative_w)}
	=
	\frac{\frac{n}{2} + \frac{1}{m}\max\{1,\, 2^{\rev{\log_2 n-m+1}}\}}
	{\frac{1}{m}\max\{1,\, 2^{\rev{\log_2 n-m+1}}\}}
	=
	\frac{mn}{2\max\{1,\, 2^{\rev{\log_2 n-m+1}}\}} + 1.
	\]
	
	This proves the lemma.
\end{proof}

Recall that in the offline setting, the best known lower bound is $\Omega(m^2)$, and this bound is tight. In contrast, \rev{within the parameter range of Lemma~\ref{lem:util-det-adv-2}, the online lower bound $\Omega(mn)$ separates the two settings as $n/m$ grows.}

In Lemma~\ref{lem:util-det-adv-ub}, we demonstrate that the bound established in Lemma~\ref{lem:util-det-adv-2} is nearly tight. Specifically, in Lemma~\ref{lem:util-det-adv-ub}, we present an algorithm achieving a distortion of $\frac{2}{3}m(n + \frac{1}{4})$.

	\begin{algorithm}[H]
	\caption{Selection Algorithm}
	\label{alg:util-det}
	\begin{algorithmic}[1]
		\STATE $\alternative_w \gets \alternative_1$ \COMMENT{Initialize with first alternative}
		\FOR{$t \gets 2$ \textbf{to} \rev{$m$}}
		\STATE Let $\alternative_t$ be the incoming alternative
		\IF {$\plu{t}{\alternative_w} = 0$ \textbf{or} $\plu{{t}}{\alternative_t} \ge \max(\plu{{t}}{\alternative_w}, n/3)$}
		\STATE $\alternative_w \gets \alternative_t$ \COMMENT{Switch selection}
		\ENDIF
		\ENDFOR
		\STATE \textbf{return} \rev{$\alternative_w$}
	\end{algorithmic}
\end{algorithm}
\begin{lemma}[Utilitarian, deterministic, adversarial order] \label{lem:util-det-adv-ub}
	\rev{For $n\ge2$,} in the online voting problem under the utilitarian assumption and adversarial arrival order of alternatives, Algorithm~\ref{alg:util-det} achieves distortion at most
	$
	\nicefrac{2}{3}m\left(n+\nicefrac{1}{4}\right).
	$
\end{lemma}
\begin{proof}
	
	The algorithm~\ref{alg:util-det} initializes by selecting $\alternative_1$ as the current winner. Subsequently, at each step $t$, it updates the winner to $\alternative_t$ if either no voter ranks the current winner as their top choice (i.e., $\plu{t}{\alternative_w} = 0$), or if the new alternative $\alternative_t$ is the top choice of at least $\max(\plu{t}{\alternative_w}, n/3)$ voters. Let the final alternative selected by the Algorithm~\ref{alg:util-det} be denoted by $\alternative_{w}$ and the optimal alternative be denoted by $\alternative_{i^*}$.
	Since $\plu{{m}}{\alternative_{w}} \ge 1$, we have
	$
	\SW(\alternative_{w}) \ge \frac{1}{m}.
	$
	Furthermore, we have $\SW(\alternative_{i^*}) \le n$. If $\plu{{m}}{\alternative_{w}} \ge 2$, then
	$
	\SW(\alternative_{w}) \ge \frac{2}{m},
	$
	which implies that the distortion is at most $\frac{1}{2}mn$, and the proof is complete.
	Hence, we focus on the remaining case that $\plu{{m}}{\alternative_{w}} = 1$. In this case, it suffices to show that
	$$
	\SW(\alternative_{i^*}) \le \frac{2}{3}n + \frac{1}{6}.
	$$
	If $\plu{{m}}{\alternative_{i^*}} \le \frac{n}{3}$, then the desired bound holds:
	$$
	\SW(\alternative_{i^*}) \le \frac{1}{3}n \cdot 1 + \frac{2}{3}n \cdot \frac{1}{2} = \frac{2}{3}n.
	$$
	Thus, we consider the remaining case where $\plu{{m}}{\alternative_{i^*}} \rev{>} \frac{n}{3}$. This implies that for any $i \ge i^*$ we have $\plu{{i}}{\alternative_{i^*}} \ge \frac{n}{3}$.
	Assume that the algorithm does not select $\alternative_{i^*}$ in the end\rev{ (otherwise, its distortion is $1$)}.
	Let $\alternative_{i}$ denote the last alternative to which the algorithm switched
	when the condition
	$$
	\plu{{i}}{\alternative_i} \ge \max\bigl(\plu{{i}}{\alternative_w},\, n/3\bigr)
	$$
	was satisfied. \rev{If there was no such switch, set $i=1$. After step $i$, the algorithm switches only when its current winner has no first-place supporters. Thus $\plu{m}{\alternative_i}\le1$: either $\alternative_i$ is the final winner, or it was discarded with zero support. Also $i\ne i^*$, since $\alternative_{i^*}$ has positive final support and could not be discarded by such a switch.} We distinguish two cases depending on the arrival order of $\alternative_{i}$ and $\alternative_{i^*}$.
	\paragraph{\textbf{Case 1}: \(i > i^*\).}
	
	Define the following two sets of voters:
	\[
	X = \voters^{i-1}(\alternative_{i^*}) \cap \voters^{i}(\alternative_i), 
	\qquad
	Y = \left(\voters \setminus \voters^{i-1}(\alternative_{i^*})\right) \cap \voters^{i}(\alternative_i).
	\]
	Thus, \(X\) consists of voters who ranked \(\alternative_{i^*}\) first before step \(i\) and switch to ranking \(\alternative_i\) first at step \(i\), while \(Y\) consists of voters who newly rank \(\alternative_i\) first but did not previously rank \(\alternative_{i^*}\) first.
	
	By construction,
	$
	|X| + |Y| = |\voters^{i}(\alternative_i)| \ge \nicefrac{n}{3}.
	$
	Moreover, the number of voters who rank \(\alternative_{i^*}\) first at the end is at most those who ranked it first before step \(i\) and did not switch, i.e.,
	$
	|\voters^{m}(\alternative_{i^*})| \le |\voters^{i-1}(\alternative_{i^*}) \setminus X|.
	$
	
	We now argue about the rank of \(\alternative_{i^*}\) for voters in \(Y\). For any voter in \(Y\), \(\alternative_{i^*}\) was not ranked first before step \(i\), and hence had rank at least \(2\). When \(\alternative_i\) arrives and becomes their top choice, the rank of \(\alternative_{i^*}\) increases to at least \(3\). 
	
	Finally, since \(\plu{m}{\alternative_w} = 1\), \rev{the alternative \(\alternative_i\)} is ranked first by at most one voter at step \(m\). Therefore, for all but at most one voter in \(Y\), the rank of \(\alternative_{i^*}\) increases by at least one more position in later steps, and is thus at least \(4\).

	We now bound $\SW(\alternative_{i^*})$.
	
	\paragraph{\textbf{Case 1a}: $|\voters^{{i-1}}(\alternative_{i^*})| \le \frac{n}{2}$.}
	
	\begin{align*}
		\SW(\alternative_{i^*})
		&\le |\voters^{i-1}(\alternative_{i^*})\setminus X|
		+ (|Y|-1)\cdot\frac14+\frac13\\
		&\quad +|\voters\setminus(Y\cup(\voters^{i-1}(\alternative_{i^*})\setminus X))|\cdot\frac12\\
		&=\frac12|\voters^{i-1}(\alternative_{i^*})|+\frac n2-\frac{|X|}{2}-\frac{|Y|}{4}+\frac1{12}\\
		&\rev{\le}\frac n4+\frac n2-\frac{|X|}{2}-\frac{|Y|}{4}+\frac1{12}\\
		&\le\frac n4+\frac n2-\frac n{12}+\frac1{12}
		=\frac{2n}{3}+\frac1{12}.
	\end{align*}
	\paragraph{\textbf{Case 1b}: \( |\voters^{i-1}(\alternative_{i^*})| > \frac{n}{2} \).}	
	
	In this case, \(\alternative_{i^*}\) is the current winner immediately before the arrival of \(\alternative_i\), and the algorithm switches to \(\alternative_i\) at step \(i\). \rev{Indeed, $\alternative_{i^*}$ had a strict majority when it arrived, and could not have been replaced while keeping that majority.} Let
	\[
	k = |\voters^{i}(\alternative_i) \cap \voters^{i-1}(\alternative_{i^*})|
	\]
	denote the number of voters who previously ranked \(\alternative_{i^*}\) first and switch to ranking \(\alternative_i\) first at step \(i\). In particular, \(\alternative_i\) is ranked first by at least \(k\) voters.
	
	Since \(\plu{m}{\alternative_w} = 1\), at most one voter ranks \(\alternative_i\) first at the final step. Therefore, among the \(k\) voters who rank \(\alternative_i\) first at step \(i\), at most one continues to rank it first at step \(m\). For the remaining \(k-1\) voters, the rank of \(\alternative_{i^*}\) is at least \(3\), and thus each contributes at most \(1/3\) to \(\SW(\alternative_{i^*})\). The one exceptional voter may contribute up to \(1/2\).
	
	\begin{revision}
		In addition, let $r=|\voters^{i-1}(\alternative_{i^*})\setminus X|$ and $z=n-r-k-|Y|$. The switch condition gives $r\le k+|Y|$. At most $r$ voters still rank $\alternative_{i^*}$ first. Voters in $Y$ contribute at most $1/4$ each, apart from the possible exceptional voter described above; the remaining $z$ voters contribute at most $1/2$ each. There is at most one exceptional voter in $X\cup Y$, and an additive $1/6$ covers either possibility. Combining these bounds, we obtain
		\begin{align*}
			\SW(\alternative_{i^*})
			&\le r+\frac{k}{3}+\frac{|Y|}{4}+\frac{z}{2}+\frac16\\
			&=\frac{2n}{3}+\frac{r-k-|Y|}{3}-\frac{|Y|}{12}-\frac{z}{6}+\frac16\\
			&\le\frac{2n}{3}+\frac16.
		\end{align*}
	\end{revision}
	
	\paragraph{\textbf{Case 2}: $i< i^*$.}
	\rev{The incumbent immediately before $i^*$ is still $\alternative_i$. Otherwise, a switch with zero incumbent support after $i$ must have selected an alternative with fewer than $n/3$ supporters; its support cannot increase, so the arrival of $\alternative_{i^*}$ would force another threshold switch, contradicting the definition of $i$.}
	Since the algorithm does not switch from $\alternative_{i}$ to $\alternative_{i^*}$, and $\plu{{m}}{\alternative_{i^*}} \ge \frac{n}{3}$, it follows that,
	$
	\plu{{{i^*}}}{\alternative_{i}} \ge \plu{{{i^*}}}{\alternative_{i^*}}.
	$
	
	We now show that at the end, for all but at most one voter in $\voters^{\rev{i^*}}(\alternative_i)$, the rank of $\alternative_{i^*}$ is at least $3$.  
	At the arrival of $\alternative_{i^*}$, voters in $\voters^{{i^*}}(\alternative_i)$ do not rank $\alternative_{i^*}$ first, thus its rank is at least $2$. Since $\voters^{\rev{m}}(\alternative_i)$ has at most one voter, for all but at most one voter the final rank of $\alternative_{i^*}$ is at least $3$.
	
	We can now bound the social welfare:
	\begin{align*}
		\SW(\alternative_{i^*})
		&\le |\voters^{i^*}(\alternative_{i^*})|
		+\bigl(|\voters^{i^*}(\alternative_i)|-1\bigr)\cdot\frac13+\frac12\\
		&\quad +|\voters\setminus(\voters^{i^*}(\alternative_i)\cup\voters^{i^*}(\alternative_{i^*}))|\cdot\frac12.
	\end{align*}
	This simplifies to
	\[
	\SW(\alternative_{i^*}) \rev{\le} \frac{1}{2}|\voters^{{i^*}}(\alternative_{i^*})| + \frac{1}{2}n - \frac{1}{6}|\voters^{{i^*}}(\alternative_i)| + \frac{1}{6}.
	\]
	Since $|\voters^{{i^*}}(\alternative_i)| \ge |\voters^{{i^*}}(\alternative_{i^*})|$, we obtain
	\begin{align*}
		\SW(\alternative_{i^*}) &\le \frac{1}{3}|\voters^{{i^*}}(\alternative_{i^*})| + \frac{1}{2}n + \frac{1}{6} \\
		&\le \frac{1}{3}\left(\frac{n}{2}\right) + \frac{1}{2}n + \frac{1}{6} = \frac{2}{3}n + \frac{1}{6}.
	\end{align*}
	
	\rev{Dividing by $\SW(\alternative_w)\ge1/m$ gives distortion at most $m(2n/3+1/6)=\frac23m(n+\frac14)$.} This completes the proof.
\end{proof}

\subsection{Randomized Rules}
Although the bound in Lemma~\ref{lem:util-det-adv-2} is almost tight \rev{within its stated parameter range}, it is significantly worse than what is achievable in the offline setting. Recall that in the offline model, the plurality voting rule achieves a distortion of \(O(m^2)\) \cite{caragiannis2011voting}. In contrast, the online setting with deterministic rules can incur substantially higher distortion.
Allowing randomness in the algorithm, however, significantly reduces this gap between the offline and online settings. The key ingredient for this improvement is Lemma~\ref{lem:gen:random}. 
	
Before stating Lemma~\ref{lem:gen:random}, we note that although the arrival order of alternatives is adversarial, we assume it is non-adaptive. That is, \rev{the preference profile and the order are fixed in advance} and do not depend on the execution of the voting rule or on its internal randomness. The voting process is then carried out with respect to this fixed order. This assumption is crucial for the correctness and analysis of our randomized rule, since different realizations of the randomness are evaluated under the same arrival sequence. A similar non-adaptive (oblivious adversary) assumption is standard in related online and streaming models, although stronger adaptive adversary models are also studied in some settings.

\begin{definition}[\rev{Regular voting rule}]\label{def:monotonicity}\label{def:regularity}
	Let \(\mathcal{E} = (\mathcal{V}, \mathcal{A}, \rev{\boldsymbol{\ranklist}})\) be an election instance, and let \(\mathcal{E}^+ = (\mathcal{V}, \mathcal{A} \cup \{\alternative^+\}, \rev{\boldsymbol{\ranklist}^{+}})\) be an instance obtained by adding a new alternative \(\alternative^+\), \rev{without changing any voter's relative ranking of the existing alternatives}. 
	An offline voting rule \(\votingrule\) is \emph{\rev{regular}} if, for every alternative \(\alternative \in \mathcal{A}\), the probability that \(\votingrule\) selects \(\alternative\) in \(\mathcal{E}\) is at least as large as the probability that it selects \(\alternative\) in \(\mathcal{E}^+\).
\end{definition}

\rev{Regularity} is a natural property of voting rules, stating that adding a new alternative should not increase the probability of selecting any existing alternative. 
In Lemma~\ref{lem:gen:random}, we show that this property enables the transformation of offline randomized voting rules into online ones. The lemma applies to both the utilitarian and metric frameworks.
\begin{lemma}[Randomized, Utilitarian and Metric, adversarial order]\label{lem:gen:random}
Let $\votingrule$ be a \rev{regular, order-independent ordinal} randomized voting rule for offline voting, \rev{defined on every nonempty finite set of alternatives}. Then, there exists an online randomized voting rule \rev{with the same outcome distribution as $\votingrule$ at every prefix, and hence} with the same distortion as $\votingrule$.
\end{lemma}
\begin{proof}
	Let $\mathcal{E} = (\mathcal{V}, \mathcal{A}, \rev{\boldsymbol{\ranklist}})$ be an online election and let $\mathcal{E}^t$ be the election instance corresponding to step $t$ of the online voting problem, \rev{with each ranking restricted to the first $t$ alternatives}. Furthermore, let $p_{i,t}$ for $1 \leq i \leq t \leq m$ be the probability that the offline voting rule $\votingrule$ chooses alternative $\alternative_i$ when applied to the offline election $\mathcal{E}^t$. \rev{Conditional on the current winner being $\alternative_i$}, we define $\votingrule'$ to be an online voting rule obtained by choosing $\alternative_t$ at step $t$ with probability $1 - \frac{p_{i,t}}{p_{i,t-1}}$. This value \rev{lies in $[0,1]$ by regularity. If $p_{i,t-1}=0$, the transition can be defined arbitrarily, since this event has probability zero under the induction hypothesis}.
	
	Now, we use induction to show that after each step $t$, the probability that alternative $\alternative_i$ for $i \leq t$ is selected as the winner up to step $t$ is equal to $p_{i,t}$.
	At step $1$, the online rule $\votingrule'$ selects $\alternative_1$ as the winner. By definition of $p_{i,t}$, we have $p_{1,1} = 1$, therefore the claim holds.
	Assume that after step $t-1$, the probability that $\votingrule'$ selects $\alternative_i$ (for $i \leq t-1$) as the winner is $p_{i,t-1}$. At step $t$, \rev{conditional on the current winner being $\alternative_i$ with $p_{i,t-1}>0$}, the online rule $\votingrule'$, with probability $1 - \frac{p_{i,t}}{p_{i,t-1}}$, selects $\alternative_t$ as the new winner, and with probability $\frac{p_{i,t}}{p_{i,t-1}}$, it keeps the previous winner. For any alternative $\alternative_i$ with $i < t$ \rev{and $p_{i,t-1}>0$}, the probability that $\alternative_i$ is selected at step $t$ is:
	\begin{align*}
		\Pr[\text{$\alternative_i$ selected at step $t$}]
		&=\Pr[\text{$\alternative_i$ selected at step $t-1$}]
		\cdot\frac{p_{i,t}}{p_{i,t-1}}\\
		&=p_{i,t-1}\cdot\frac{p_{i,t}}{p_{i,t-1}}=p_{i,t}.
	\end{align*}
	\rev{If $p_{i,t-1}=0$, regularity gives $p_{i,t}=0$, and the same conclusion holds.}
	
	Note that the probability that $\votingrule'$ selects $\alternative_t$ at step $t$ is $1-\sum_{1 \leq i <t} p_{i,t} = p_{t,t}$. \rev{At the final step, equality of the outcome distributions gives the same expected social welfare or social cost on each fixed instance, and therefore the same distortion.} This completes the proof.
\end{proof}

Now, using Lemma~\ref{lem:gen:random}, for the utilitarian setting we show that distortion  $\rev{O(\sqrt{m\log m})}$ can be achieved by extending the Harmonic rule. Introduced by Boutilier et al.~\cite{boutilier2015optimal}, this rule assigns a score to each alternative based on its positions in the voters’ rankings: an alternative receives a contribution of \(1\) from a voter if it is ranked first, \(\nicefrac{1}{2}\) if it is ranked second, \(\nicefrac{1}{3}\) if it is ranked third, and so on.
The rule then selects an alternative using a simple randomized procedure: with probability \(\nicefrac{1}{2}\), it chooses an alternative uniformly at random, and with probability \(\nicefrac{1}{2}\), it selects an alternative with probability proportional to its score. Boutilier et al.~\cite{boutilier2015optimal} show that this rule has distortion \(\rev{O(\sqrt{m\log m})}\). In Lemma~\ref{lem:util-harmonic}, we show that this rule satisfies \rev{regularity}.

\begin{lemma}\label{lem:util-harmonic}
	The Harmonic voting rule is \rev{regular}.
\end{lemma}
\begin{proof}
	In the Harmonic rule, each alternative \(a\) is assigned a score
	\[
	S(a) = \sum_{v_i \in \mathcal{V}} \frac{1}{r_i(a)},
	\]
	where \(r_i(a)\) denotes the rank of alternative \(a\) in voter \(v_i\)'s ranking. The rule selects an alternative by mixing two distributions: with probability \(\nicefrac{1}{2}\), it selects an alternative uniformly at random, and with probability \(\nicefrac{1}{2}\), it selects an alternative with probability proportional to its score.
	
	Consider an instance \(\mathcal{E}\) and a second instance \(\mathcal{E}^+\) obtained by adding a new alternative \(a^+\). Fix any existing alternative \(a\). When \(a^+\) is added, the rank of \(a\) in each voter’s ranking can only weakly increase (i.e., become worse). Hence, for every voter \(v_i\),
	\[
	\frac{1}{r_i^+(a)} \le \frac{1}{r_i(a)},
	\]
	and therefore
	\[
	S^+(a) \le S(a).
	\]
	
	We now compare selection probabilities in the two parts of the rule.
	In the uniform part, the probability of selecting \(a\) decreases from \(\frac{1}{2}\cdot\frac{1}{m}\) in \(\mathcal{E}\) to \(\frac{1}{2}\cdot\frac{1}{m+1}\) in \(\mathcal{E}^+\).
	In the score-proportional part, since \(S^+(a) \le S(a)\) while total score increases after adding \(a^+\) \rev{from $nH_m$ to $nH_{m+1}$, where $H_m=\sum_{j=1}^m1/j$}, the probability of selecting \(a\) in this part also weakly decreases:
	\[
	\frac{S^+(a)}{\sum_{b} S^+(b)} \le \frac{S(a)}{\sum_{b} S(b)}.
	\]
	
	Combining both parts, the overall probability of selecting \(a\) in \(\mathcal{E}^+\) is at most its probability in \(\mathcal{E}\). This completes the proof.
	
\end{proof}

\rev{Combining the offline distortion bound of $O(\sqrt{m\log m})$ for Harmonic~\cite{boutilier2015optimal} with Lemmas~\ref{lem:gen:random} and~\ref{lem:util-harmonic} yields Corollary~\ref{cor:util-rand-adv}.}

\begin{corollary}[Utilitarian, randomized, adversarial order] \label{cor:util-rand-adv}
	For the online voting problem under the utilitarian assumption and adversarial arrival order of alternatives, there exists a randomized algorithm whose distortion is at most
	$
	\rev{O(\sqrt{m\log m})}.
	$
\end{corollary}

Given Corollary~\ref{cor:util-rand-adv}, a natural question is whether more sophisticated voting rules can achieve a better distortion guarantee. In the offline setting, the \rev{optimal asymptotic} distortion in the utilitarian framework is $O(\sqrt{m})$ \cite{ebadian2024optimized}, achieved by a voting rule based on the notion of stable lotteries.

\begin{definition}[Stable Lottery]
For an \rev{integer} parameter $k$ with $1 \le k \le m$, a \emph{stable lottery} is a probability distribution $\mathcal{X}$ over subsets of size $k$ such that for every alternative $i \in \alternatives$,
	$
	\mathbb{E}_{X \sim \mathcal{X}}[\rev{|\voters_{i \succ X}|}] < \frac{n}{k}.
	$
\rev{Here $\voters_{i\succ X}$ is the set of voters who prefer alternative $i$ to every member of $X$.}
\end{definition}

It is shown by Cheng et al.~\cite{cheng2020group} that, for every \(k\), a stable lottery over committees of size \(k\) always exists. Ebadian et al.~\cite{ebadian2024optimized} propose a voting rule based on stable lotteries that guarantees a distortion of \(O(\sqrt{m})\). The structure of their rule is as follows: they first compute a stable lottery \(\mathcal{X}\) over committees of size \rev{$k=\lceil\sqrt{m}\rceil$}. Then, with probability \(\frac{1}{2}\), they sample a committee from \(\mathcal{X}\) and select a uniformly random member of that committee. With probability \(\frac{1}{2}\), they instead select an alternative uniformly at random.
In Lemma~\ref{lem:stable_lottery} we show that this rule is not necessarily \rev{regular}. 

\begin{lemma}\label{lem:stable_lottery}
	The stable lottery rule is not necessarily \rev{regular}. 
\end{lemma}
\begin{proof}
	We prove the claim by providing a counterexample. Consider a profile with $n$ voters (where $n$ is an even integer) and 9 alternatives with the preference profile illustrated below:

	\[
	\begin{array}{c|c|c}
		& \voter_1\cdots\voter_{n/2} & \voter_{n/2 + 1}\cdots\voter_{n} \\ \hline
		
		\text{1st} & \alternative_1 &  \alternative_2   \\
		
		\text{2nd} & \alternative_3 & \alternative_3   \\
		
		\text{3rd} & \alternative_2 & \alternative_1   \\
		
		\text{4th} & \alternative_4 & \alternative_4   \\
		
		\text{5th} & \alternative_5 & \alternative_5   \\
		
		\vdots & \vdots & \vdots \\
		
		\text{9th} & \alternative_9 & \alternative_9   
	\end{array}
	\]
	
	In step 4 we have $k = \sqrt{4} = 2$. Consider a lottery that selects the committee $\{a_1, a_2\}$ with probability $0.2$ and the committee $\{a_2, a_3\}$ with probability $0.8$. \rev{This distribution is stable: the expected number of voters preferring $a_1$ to every committee member is $0.8\cdot n/2=2n/5<n/2$, and the corresponding number for every other alternative is zero. Including the uniform half of the rule, the probability of selecting $a_1$ is
		\[
		\frac12\left(0.2\cdot\frac12\right)+\frac12\cdot\frac14=\frac7{40}.
		\]}
	
	In step 9 we have $k = \sqrt{9} = 3$. \rev{Choose the committee $\{a_1,a_2,a_3\}$ with probability one. This is a stable lottery, since the committee contains every voter's top alternative. The probability of selecting $a_1$ under the full rule is now
		\[
		\frac12\cdot\frac13+\frac12\cdot\frac19=\frac29>\frac7{40}.
		\]
		Thus the selection probability of $a_1$ must increase at some one-alternative extension between steps 4 and 9, regardless of the stable lotteries chosen at the intermediate steps. The rule is therefore not regular.}
\end{proof}

We note that the instance used in the proof of Lemma~\ref{lem:stable_lottery} admits multiple stable lotteries. \rev{Consequently, the example shows that an arbitrary choice of stable lotteries need not satisfy regularity. It does not rule out a selection of stable lotteries that satisfies regularity on every profile.}
	\section{Metric Framework}
 \label{sec:metric}
In this section, we consider the metric framework and derive upper and lower bounds on the distortion for online voting, for both deterministic and randomized mechanisms. Recall that in the offline case, deterministic voting rules achieve a tight bound of $3$ \rev{\cite{anshelevich2015approximating,gkatzelis2020resolving,kizilkaya2022pluralityveto}}, while randomized rules have a lower bound of \rev{$2.1126$~\cite{charikar2022metric}} and an upper bound of \rev{$2.13713$~\cite{cai2026stable}}.
We start by showing that a deterministic algorithm can achieve a distortion of $2+\sqrt{5}$. The underlying concept of this result is based on the definition of the weighted uncovered set \cite{Munagala2019Improved}.

\rev{For alternatives $\alternative_i$ and $\alternative_k$, let $\voters_{i\succ k}$ denote the set of voters who prefer $\alternative_i$ to $\alternative_k$.}

\begin{definition}[$\lambda$-weighted uncovered set]\label{def:lambda-weighted uncovered set}
	Let $\lambda \in [0.5,1]$ be a constant. The $\lambda$-weighted uncovered set is the set of alternatives $\alternative_i \in \alternatives$ such that for any alternative $\alternative_k \ne \alternative_i$, we have:
	\begin{itemize}
		\item either $|\voters_{i \succ k}| \ge (1-\lambda)\rev{n}$
		\item or there is another alternative $\alternative_l$ such that $|\voters_{i \succ l}| \ge (1-\lambda)\rev{n}$ and $\rev{|\voters_{l \succ k}|} \ge \lambda\rev{n}$.
	\end{itemize}
\end{definition}

Lemma~\ref{lem:metric-det-adv} builds on the work of Munagala and Wang \cite{Munagala2019Improved}, which shows that any alternative from the weighted uncovered set with weight $\lambda = \varphi = \frac{\sqrt{5} - 1}{2}$ achieves a distortion of $2+\sqrt{5}$.

\begin{lemma}[Metric, deterministic, adversarial order] \label{lem:metric-det-adv}
	For the online voting problem under the metric assumption and adversarial arrival order of alternatives, there exists a deterministic algorithm whose distortion is at most
	$
	2+\sqrt{5}.
	$
\end{lemma}

\begin{proof}
	By Theorem~3.4 of \cite{Munagala2019Improved}, every alternative in the $\varphi$-weighted uncovered set  has metric distortion at most $2+\sqrt{5}$.
	We construct an online algorithm that, after each arrival, maintains an alternative belonging to the $\varphi$-weighted uncovered set of the alternatives seen so far. The initial alternative trivially satisfies this.
	
	Assume inductively that after $t-1$ steps we hold an alternative $\alternative_w$ in the $\varphi$-weighted uncovered set of $\{\alternative_1,\dots,\alternative_{t-1}\}$.
	When $\alternative_t$ appears, define
	\[
	S = \{\, \alternative_i \mid \rev{1\le i<t,}\ i \neq w, \ |\voters_{w \succ i}| \ge (1-\varphi)\rev{n} \,\}.
	\]
	We distinguish three cases. \rev{In the first two cases we retain $\alternative_w$; in the third we replace it with $\alternative_t$.}
	
	\begin{enumerate}
		\item \textbf{$|\voters_{w \succ t}| \ge (1-\varphi)\rev{n}$.}
		The \rev{first condition} of Definition~\ref{def:lambda-weighted uncovered set} holds for the pair $(\alternative_w, \alternative_t)$; conditions for all other pairs are unchanged. Hence $\alternative_w$ remains in the $\varphi$-weighted uncovered set.
		
		\item There exists $\alternative_i \in S$ with $|\voters_{i \succ t}| \ge \varphi\rev{n}$.
		By definition of $S$, $|\voters_{w \succ i}| \ge (1-\varphi)\rev{n}$. Together with $|\voters_{i \succ t}| \ge \varphi\rev{n}$ this fulfills the \rev{second condition} for the pair $(w, \alternative_t)$ via $\alternative_i$. All other pairs are unaffected. Hence $\alternative_w$ remains in the $\varphi$-weighted uncovered set.
		
		\item Otherwise -- i.e., $|\voters_{w \succ t}| < (1-\varphi)\rev{n}$ and for every $\rev{\alternative_i} \in S$, $|\voters_{i \succ t}| < \varphi\rev{n}$.
		From $|\voters_{w \succ t}| < (1-\varphi)\rev{n}$ we obtain
		\[
		|\voters_{t \succ w}| = \rev{n} - |\voters_{w \succ t}| > \varphi\rev{n} \ge (1-\varphi)\rev{n},
		\]
		so the \rev{first condition} holds for $(\alternative_t, w)$.
		For any $\alternative_i \in S$, $|\voters_{i \succ t}| < \varphi\rev{n}$ gives
		$
		|\voters_{t \succ i}| = \rev{n} - |\voters_{i \succ t}| > (1-\varphi)\rev{n},
		$
		and therefore, the \rev{first condition} holds for $(\alternative_t, \alternative_i)$ as well.
		Now consider any other \rev{previously arrived} alternative $\alternative_l \notin S \cup \{\alternative_t,\alternative_w\}$. Since $w$ was in the uncovered set before step $t$, there must exist some $\alternative_i \in S$ with $|\voters_{i \succ l}| \ge \varphi\rev{n}$; otherwise the \rev{first condition} for $(\alternative_w,\alternative_l)$ would force $\alternative_l \in S$. Together with the already established $|\voters_{t \succ i}| \ge (1-\varphi)\rev{n}$ this satisfies the \rev{second condition} for $(\alternative_t, \alternative_l)$ via $\alternative_i$.
		Hence $\alternative_t$ itself satisfies all requirements of the $\varphi$-weighted uncovered set.
	\end{enumerate}
	
	In every case we obtain a current alternative that lies in the $\varphi$-weighted uncovered set of $\{\alternative_1,\dots,\alternative_t\}$. By induction, after the last step the algorithm outputs an alternative from this set for the whole instance, guaranteeing distortion at most $2+\sqrt{5}$.
\end{proof}

Recall that in the metric framework, the best possible \rev{offline deterministic} distortion is $3$, and this tight bound is achieved by matching-based voting rules such as plurality matching and plurality veto (see \cite{gkatzelis2020resolving}\rev{, \cite{kizilkaya2022pluralityveto}}). In Lemma~\ref{lem:metric-det-adv}, we established an upper bound of $2+\sqrt{5}$ for the online setting.
A natural question then arises: \emph{Can these matching-based rules---which attain distortion $3$ offline---be adapted to work online?} 

The main issue is that these rules require checking whether a perfect matching exists in the domination graph of a given alternative. \rev{Although this check can be performed on the alternatives seen so far, the arrival of a new alternative can change which domination graphs admit a perfect matching. An alternative needed at the final step may already have been irrevocably rejected.}

\begin{definition}[Domination graph]\label{def:domination-graph}
	For a voter $\voter_j$, let $top(\voter_j)$ denote the highest-ranked alternative in $\ranklist_j$.  
	Given an election $\mathcal{E} = (\mathcal{V}, \mathcal{A}, \rev{\boldsymbol{\ranklist}})$ and an alternative $\alternative_i \in \mathcal{A}$, the \emph{domination graph} of $\alternative_i$ is the bipartite graph \rev{on two copies of the voter set} $G(\alternative_i) = (\mathcal{V}, \mathcal{V}, E_i)$, where $(\voter, \voter') \in E_i$ if and only if $\alternative_i \succeq_{\voter} top(\voter')$.
\end{definition}

Gkatzelis et al.~\cite{gkatzelis2020resolving} show that in every election instance, there exists an alternative whose domination graph admits a perfect matching. It is further shown that any alternative with this property guarantees distortion at most \(3\). Matching-based voting rules are built on identifying such an alternative. \rev{The next example shows why such an alternative cannot always be retained online.}

However, to see why such rules cannot be implemented in the online setting, consider Table~\ref{table:matching-counterexample}. The two instances in Table~\ref{table:matching-counterexample-a} and Table~\ref{table:matching-counterexample-b} are identical up to step~3, that is, before the arrival of $\alternative_4$. Hence, at that point, any deterministic online rule observes exactly the same information in both instances and must make the same decision.

However, once $\alternative_4$ is revealed, the structure changes. In Table~\ref{table:matching-counterexample-a}, \rev{only} the domination graph of $\alternative_1$ admits a perfect matching, whereas in Table~\ref{table:matching-counterexample-b}, \rev{only the domination graph of $\alternative_3$ admits} a perfect matching (see Figures~\ref{fig:dg1} and~\ref{fig:dg2}).
Thus, the identity of the alternative with a perfect matching depends on information that is not available before step~4. \rev{Let $\alternative_w$ be the winner retained after step~3. If $\alternative_w\ne\alternative_1$, the rule fails on the first instance: at step~4 it can select only $\alternative_w$ or $\alternative_4$, neither of which has a perfect matching. If $\alternative_w=\alternative_1$, it fails on the second instance for the same reason, since only $\alternative_3$ has a perfect matching there.} Since both instances are indistinguishable at step~3, no deterministic online rule can guarantee selecting an alternative whose domination graph admits a perfect matching.
Therefore,  Lemma~\ref{lemma:metric-deterministic-no-mathcing} holds.

\begin{table}[htbp]
	\centering
	\begin{subtable}[t]{0.45\textwidth}
		\centering
		\begin{tabular}[t]{c|c|c|c|c}
			& $\voter_1$ & $\voter_2$ & $\voter_3$ & $\voter_4$ \\ \hline
			\text{1st} & $\alternative_1$  & $\alternative_1$ & $\alternative_4$ & $\alternative_3$  \\
			\text{2nd} & $\alternative_4$  & $\alternative_4$ & $\alternative_3$ & $\alternative_2$  \\
			\text{3rd} & $\alternative_2$  & $\alternative_2$ & $\alternative_2$ & $\alternative_1$  \\
			\text{4th} & $\alternative_3$  & $\alternative_3$ & $\alternative_1$ & $\alternative_4$
		\end{tabular}
		\caption{}
		\label{table:matching-counterexample-a}
	\end{subtable}
	\hspace{1em}          
	\begin{subtable}[t]{0.45\textwidth}
		\centering
		\begin{tabular}[t]{c|c|c|c|c}
			& $\voter_1$ & $\voter_2$ & $\voter_3$ & $\voter_4$ \\ \hline
			\text{1st} & $\alternative_1$  & $\alternative_4$ & $\alternative_3$ & $\alternative_3$  \\
			\text{2nd} & $\alternative_2$  & $\alternative_1$ & $\alternative_4$ & $\alternative_4$  \\
			\text{3rd} & $\alternative_3$  & $\alternative_2$ & $\alternative_2$ & $\alternative_2$  \\
			\text{4th} & $\alternative_4$  & $\alternative_3$ & $\alternative_1$ & $\alternative_1$
		\end{tabular}
		\caption{}
		\label{table:matching-counterexample-b}
	\end{subtable}
	\caption{In Table~\ref{table:matching-counterexample-a}, only the domination graph of $\alternative_1$ admits a perfect matching, while in Table~\ref{table:matching-counterexample-b}, only the domination graph of $\alternative_3$ admits a perfect matching.}
	\label{table:matching-counterexample}
\end{table}
\tikzset{
	every node/.style={circle, draw, minimum size=0.6cm, inner sep=0pt, font=\small},
	normaledge/.style={black, thin},
	matching/.style={red, thick}
}

\begin{figure}[!t] 
	\centering

	\vspace{-1mm} 
	
	\begin{subfigure}[b]{0.46\columnwidth}
		\centering
		\begin{tikzpicture}[scale=0.67, every node/.style={transform shape, draw, circle, inner sep=1pt}]
			\node (L1) at (0,0) {\(v_1\)}; \node (L2) at (0,-1) {\(v_2\)}; \node (L3) at (0,-2) {\(v_3\)}; \node (L4) at (0,-3) {\(v_4\)};
			\node (R1) at (1.5,0) {\(v_1\)}; \node (R2) at (1.5,-1) {\(v_2\)}; \node (R3) at (1.5,-2) {\(v_3\)}; \node (R4) at (1.5,-3) {\(v_4\)};
			\foreach \i in {1,2} { \foreach \j in {1,2,3,4} { \draw[gray, very thin] (L\i) -- (R\j); } }
			\draw[gray, very thin] (L3) -- (R1); \draw[gray, very thin] (L3) -- (R2);
			\draw[gray, very thin] (L4) -- (R1); \draw[gray, very thin] (L4) -- (R2); \draw[gray, very thin] (L4) -- (R3);
			\draw[matching] (L1) -- (R3); \draw[matching] (L2) -- (R4); \draw[matching] (L3) -- (R1); \draw[matching] (L4) -- (R2);
		\end{tikzpicture}
		\caption{\(a_1\)}
	\end{subfigure}
	\hfill
	\begin{subfigure}[b]{0.46\columnwidth}
		\centering
		\begin{tikzpicture}[scale=0.67, every node/.style={transform shape, draw, circle, inner sep=1pt}]
			\node (L1) at (0,0) {\(v_1\)}; \node (L2) at (0,-1) {\(v_2\)}; \node (L3) at (0,-2) {\(v_3\)}; \node (L4) at (0,-3) {\(v_4\)};
			\node (R1) at (1.5,0) {\(v_1\)}; \node (R2) at (1.5,-1) {\(v_2\)}; \node (R3) at (1.5,-2) {\(v_3\)}; \node (R4) at (1.5,-3) {\(v_4\)};
			\draw (L1) -- (R4); \draw (L2) -- (R4);
			\foreach \j in {1,2} { \draw (L3) -- (R\j); }
			\foreach \j in {1,2,3} { \draw (L4) -- (R\j); }
		\end{tikzpicture}
		\caption{\(a_2\) }
	\end{subfigure}
	\par\medskip
	\begin{subfigure}[b]{0.46\columnwidth}
		\centering
		\begin{tikzpicture}[scale=0.67, every node/.style={transform shape, draw, circle, inner sep=1pt}]
			\node (L1) at (0,0) {\(v_1\)}; \node (L2) at (0,-1) {\(v_2\)}; \node (L3) at (0,-2) {\(v_3\)}; \node (L4) at (0,-3) {\(v_4\)};
			\node (R1) at (1.5,0) {\(v_1\)}; \node (R2) at (1.5,-1) {\(v_2\)}; \node (R3) at (1.5,-2) {\(v_3\)}; \node (R4) at (1.5,-3) {\(v_4\)};
			\draw (L1) -- (R4); \draw (L2) -- (R4);
			\foreach \j in {1,2,4} { \draw (L3) -- (R\j); }
			\foreach \j in {1,2,3,4} { \draw (L4) -- (R\j); }
		\end{tikzpicture}
		\caption{\(a_3\)}
	\end{subfigure}
	\hfill
	\begin{subfigure}[b]{0.46\columnwidth}
		\centering
		\begin{tikzpicture}[scale=0.67, every node/.style={transform shape, draw, circle, inner sep=1pt}]
			\node (L1) at (0,0) {\(v_1\)}; \node (L2) at (0,-1) {\(v_2\)}; \node (L3) at (0,-2) {\(v_3\)}; \node (L4) at (0,-3) {\(v_4\)};
			\node (R1) at (1.5,0) {\(v_1\)}; \node (R2) at (1.5,-1) {\(v_2\)}; \node (R3) at (1.5,-2) {\(v_3\)}; \node (R4) at (1.5,-3) {\(v_4\)};
			\draw (L1) -- (R3); \draw (L1) -- (R4); \draw (L2) -- (R3); \draw (L2) -- (R4);
			\foreach \j in {1,2,3,4} { \draw (L3) -- (R\j); }
			\draw (L4) -- (R3);
		\end{tikzpicture}
		\caption{\(a_4\)}
	\end{subfigure}
	\caption{Domination graphs for instance of Table~\ref{table:matching-counterexample-a}}
	\label{fig:dg1}
	\vspace{2mm} 
\vspace{2mm}
 
	\centering

	\vspace{-1mm} 
	
	\begin{subfigure}[b]{0.46\columnwidth}
		\centering
		\begin{tikzpicture}[scale=0.67, every node/.style={transform shape, draw, circle, inner sep=1pt}]
			\node (L1) at (0,0) {\(v_1\)}; \node (L2) at (0,-1) {\(v_2\)}; \node (L3) at (0,-2) {\(v_3\)}; \node (L4) at (0,-3) {\(v_4\)};
			\node (R1) at (1.5,0) {\(v_1\)}; \node (R2) at (1.5,-1) {\(v_2\)}; \node (R3) at (1.5,-2) {\(v_3\)}; \node (R4) at (1.5,-3) {\(v_4\)};
			\foreach \j in {1,2,3,4} { \draw (L1) -- (R\j); }
			\foreach \j in {1,3,4} { \draw (L2) -- (R\j); }
			\draw (L3) -- (R1); \draw (L4) -- (R1);
		\end{tikzpicture}
		\caption{\(a_1\) }
	\end{subfigure}
	\hfill
	\begin{subfigure}[b]{0.46\columnwidth}
		\centering
		\begin{tikzpicture}[scale=0.67, every node/.style={transform shape, draw, circle, inner sep=1pt}]
			\node (L1) at (0,0) {\(v_1\)}; \node (L2) at (0,-1) {\(v_2\)}; \node (L3) at (0,-2) {\(v_3\)}; \node (L4) at (0,-3) {\(v_4\)};
			\node (R1) at (1.5,0) {\(v_1\)}; \node (R2) at (1.5,-1) {\(v_2\)}; \node (R3) at (1.5,-2) {\(v_3\)}; \node (R4) at (1.5,-3) {\(v_4\)};
			\foreach \j in {2,3,4} { \draw (L1) -- (R\j); }
			\foreach \j in {3,4} { \draw (L2) -- (R\j); }
			\draw (L3) -- (R1); \draw (L4) -- (R1);
		\end{tikzpicture}
		\caption{\(a_2\)}
	\end{subfigure}
	\hfill
	\begin{subfigure}[b]{0.46\columnwidth}
		\centering
		\begin{tikzpicture}[scale=0.67, every node/.style={transform shape, draw, circle, inner sep=1pt}]
			\node (L1) at (0,0) {\(v_1\)}; \node (L2) at (0,-1) {\(v_2\)}; \node (L3) at (0,-2) {\(v_3\)}; \node (L4) at (0,-3) {\(v_4\)};
			\node (R1) at (1.5,0) {\(v_1\)}; \node (R2) at (1.5,-1) {\(v_2\)}; \node (R3) at (1.5,-2) {\(v_3\)}; \node (R4) at (1.5,-3) {\(v_4\)};
			\foreach \j in {2,3,4} { \draw[gray, very thin] (L1) -- (R\j); }
			\foreach \j in {3,4} { \draw[gray, very thin] (L2) -- (R\j); }
			\foreach \j in {1,2,3,4} { \draw[gray, very thin] (L3) -- (R\j); \draw[gray, very thin] (L4) -- (R\j); }
			\draw[matching] (L1) -- (R2); \draw[matching] (L2) -- (R3); \draw[matching] (L3) -- (R1); \draw[matching] (L4) -- (R4);
		\end{tikzpicture}
		\caption{\(a_3\)}
	\end{subfigure}
	\hfill
	\begin{subfigure}[b]{0.46\columnwidth}
		\centering
		\begin{tikzpicture}[scale=0.67, every node/.style={transform shape, draw, circle, inner sep=1pt}]
			\node (L1) at (0,0) {\(v_1\)}; \node (L2) at (0,-1) {\(v_2\)}; \node (L3) at (0,-2) {\(v_3\)}; \node (L4) at (0,-3) {\(v_4\)};
			\node (R1) at (1.5,0) {\(v_1\)}; \node (R2) at (1.5,-1) {\(v_2\)}; \node (R3) at (1.5,-2) {\(v_3\)}; \node (R4) at (1.5,-3) {\(v_4\)};
			\draw (L1) -- (R2);
			\foreach \j in {1,2,3,4} { \draw (L2) -- (R\j); }
			\foreach \j in {1,2} { \draw (L3) -- (R\j); \draw (L4) -- (R\j); }
		\end{tikzpicture}
		\caption{\(a_4\)}
	\end{subfigure}
	\caption{Domination graphs for the instance in Table~\ref{table:matching-counterexample-b}}
	\label{fig:dg2}
	\vspace{2mm} 
\end{figure}

\begin{lemma}\label{lemma:metric-deterministic-no-mathcing}
	For the online voting problem under the metric assumption and adversarial arrival order of alternatives, 
	no deterministic algorithm can guarantee a perfect matching in the domination graph of the selected alternative. 
\end{lemma}
\begin{proof}
The two instances in Table~\ref{table:matching-counterexample} have identical rankings on the first three alternatives. Thus a deterministic rule retains the same incumbent in both at that time. In the first instance only $a_1$ has a perfect matching, whereas in the second only $a_3$ does. Neither is the last arrival $a_4$, so a rule retaining one of them at time $3$ necessarily fails in the other instance.
\end{proof}

\FloatBarrier
The preceding example rules out selecting an alternative whose domination graph always admits a perfect matching. It does not by itself give a distortion lower bound greater than $3$. We now prove such a bound, which separates the online model from the optimal offline guarantee.
\begin{lemma}\label{lem:metric-det-lower}
For integers $N\ge2$ and $1\le k<N$, with $n=2N$ voters and three alternatives, every deterministic online rule has metric distortion at least
\[
 L(N,k)=\min\left\{\frac{5N-k}{N+k},\;3+\frac{2k}{N}\right\}.
\]
Consequently, its worst-case distortion over electorate sizes is at least $\sqrt{13}$. 
\end{lemma}
\begin{proof}
Fix an arbitrary deterministic online rule. Let
$n=2N$ and consider the arrival of two alternatives, $a_1$
and $a_2$, such that exactly $N$ voters prefer each
alternative. Without loss of generality, suppose that the rule retains $\alternative_2$. Let $X$ be the $N$ voters preferring $\alternative_1$ to $\alternative_2$. Partition the remaining voters into $Y$ of size $k$ and $Z$ of size $N-k$. Reveal $\alternative_3$, completing the rankings as follows:
\[
\begin{array}{c|c|c}
\text{group}&\text{size}&\text{ranking}\\\hline
X&N&\alternative_1 \succ \alternative_3 \succ \alternative_2\\
Y&k&\alternative_2 \succ \alternative_1 \succ \alternative_3\\
Z&N-k&\alternative_3 \succ \alternative_2 \succ \alternative_1
\end{array}
\]
Both initial comparisons are preserved. Since $\alternative_1$ is unavailable, the final output belongs to $\{\alternative_2,\alternative_3\}$.
We give two metrics consistent with this same complete ordinal profile which are illustrated in Figure~\ref{fig:lemma9-metrics}.

\begin{figure}[t]
	\centering
	
	\begin{subfigure}[t]{0.48\linewidth}
		\centering
		\begin{tikzpicture}[
			x=1.5cm, y=1.5cm,
			vertex/.style={
				circle, draw, fill=black,
				minimum size=1mm, inner sep=0pt,
				font=\small
			},
			every label/.style={
				draw=none, fill=none, shape=rectangle,
				inner sep=0pt, outer sep=0pt,
				font=\scriptsize
			},
			edge/.style={thick}
			]
			\node[vertex, label=below:$X$] (x) at (0,0) {$x$};
			\node[vertex, label=above:{$Z,\alternative_3$}] (z) at (0,1) {$z$};
			\node[vertex, label=above:$Y$] (y) at (1,1) {$y$};
			\node[vertex, label=below:$\alternative_1$] (p) at (1,0) {$p$};
			\node[vertex, label=above:$\alternative_2$] (q) at (2,1) {$q$};
			
			\draw[edge] (x)--(z)--(y)--(p)--(x);
			\draw[edge] (y)--(q);
		\end{tikzpicture}
		\caption{Metric 1}
		\label{fig:lemma9-metric1}
	\end{subfigure}
	\hfill
	\begin{subfigure}[t]{0.48\linewidth}
		\centering
		\begin{tikzpicture}[
			x=1.5cm, y=1.5cm,
			vertex/.style={
				circle, draw, fill=black,
				minimum size=1mm, inner sep=0pt,
				font=\small
			},
			every label/.style={
				draw=none, fill=none, shape=rectangle,
				inner sep=0pt, outer sep=0pt,
				font=\scriptsize
			},
			edge/.style={thick}
			]
			\node[vertex, label=below:{$X,\alternative_1$}] (x) at (0,0) {$x$};
			\node[vertex, label=below:$Y$] (y) at (1,0) {$y$};
			\node[vertex, label=above:$Z$] (z) at (0,1) {$z$};
			\node[vertex, label=above:$\alternative_2$] (p) at (1,1) {$p$};
			\node[vertex, label=above:$\alternative_3$] (q) at (-1,1) {$q$};
			
			\draw[edge] (x)--(y)--(p)--(z)--(x);
			\draw[edge] (z)--(q);
		\end{tikzpicture}
		\caption{Metric 2}
		\label{fig:lemma9-metric2}
	\end{subfigure}
	
	\caption{
		The two unit-edge graphs defining the metrics in Lemma~9.
	}
	\label{fig:lemma9-metrics}
\end{figure}
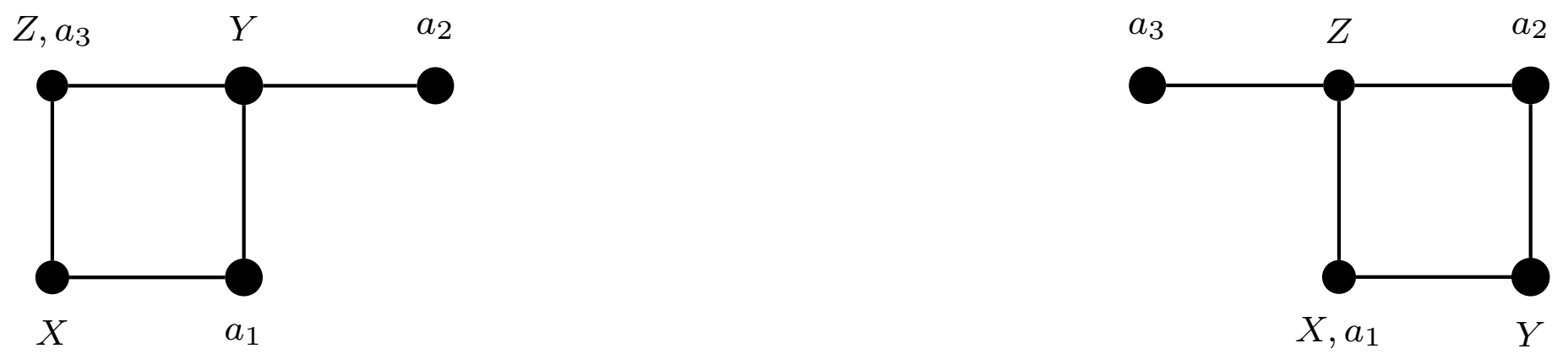

The respective voter--alternative distance tables are
\[
\begin{array}{c|rrr}
	\multicolumn{4}{c}{\text{Metric 1}}\\
	&\alternative_1&\alternative_2&\alternative_3\\\hline
	X&1&3&1\\
	Y&1&1&1\\
	Z&2&2&0
\end{array}
\qquad
\begin{array}{c|rrr}
	\multicolumn{4}{c}{\text{Metric 2}}\\
	&\alternative_1&\alternative_2&\alternative_3\\\hline
	X&0&2&2\\
	Y&1&1&3\\
	Z&1&1&1
\end{array}
\]
Shortest-path realizations guarantee all triangle inequalities, including those involving pairs of voters or alternatives. Both tables respect all the stated rankings with permissible distance ties. Their social costs, in the order $(\alternative_1,\alternative_2,\alternative_3)$, are
\begin{align*}
 (\SC_1(\alternative_1),\SC_1(\alternative_2),\SC_1(\alternative_3))
   &=(3N-k,\;5N-k,\;N+k),\\
 (\SC_2(\alternative_1),\SC_2(\alternative_2),\SC_2(\alternative_3))
   &=(N,\;3N,\;3N+2k).
\end{align*}
In the first metric, $\alternative_3$ is optimal, and selecting $\alternative_2$ gives a ratio of $(5N-k)/(N+k)$. In the second, $\alternative_1$ is optimal, and selecting $\alternative_3$ gives a ratio of $3+2k/N$. The rule sees identical rankings in both metrics and hence makes identical decisions. Selecting the appropriate metric for its output proves the bound. For a fixed deterministic rule, the construction specifies a fixed complete instance; it does not require revealing or changing distances during execution.

For $N=10,k=3$, the bound is $\min\{47/13,18/5\}=18/5$. Writing $r=k/N$, the two expressions become $(5-r)/(1+r)$ and $3+2r$. They coincide when $r^2+3r-1=0$. Rational approximations to $r=(\sqrt{13}-3)/2\in(0,1)$ therefore give lower bounds converging to $\sqrt{13}$.
\end{proof}

Together with Lemma~\ref{lem:metric-det-adv}, this places the optimal uniform deterministic online metric guarantee in $[\sqrt{13},2+\sqrt5]$. The lower endpoint is a supremum over electorate sizes, not a claim of the same bound for every fixed $n$.

We now give a deterministic online rule with distortion at most $\sqrt{13}$ for every number of voters when there are three alternatives. Together with Lemma~\ref{lem:metric-det-lower}, this establishes the optimal uniform bound. 

\begin{lemma}\label{lem:three-comparison}
	Consider an election with three alternatives $(\alternative_1,\alternative_2,\alternative_3)$. Let
	\[
	\begin{aligned}
		x_1&=|\voters_{\alternative_1 \succ \alternative_2}|,\\
		x_2&=|\voters_{\alternative_2 \succ \alternative_1 \succ \alternative_3}|,\\
		x_3&=|\voters_{\alternative_2 \succ \alternative_3 \succ \alternative_1}|,\\
		x_4&=|\voters_{\alternative_3 \succ \alternative_2 \succ \alternative_1}|,
	\end{aligned}
	\]
	so $x_1+x_2+x_3+x_4=n$. For any $\tau\ge0$, if
	\[
	x_2+x_3+x_4\le\tau(x_1+x_4),\qquad
	x_3+x_4\le\tau(x_1+x_2),
	\]
	then every consistent metric satisfies
	$\SC(\alternative_1)\le(1+2\tau)\SC(\alternative_2)$.
\end{lemma}

\begin{proof}
	By the triangle inequality, each voter $i$ satisfies
	$d(i,\alternative_1)-d(i,\alternative_2)
	\le d(\alternative_1,\alternative_2)$.
	Voters in $\voters_{\alternative_1 \succ \alternative_2}$ have distance to
	$\alternative_2$ at least $d(\alternative_1,\alternative_2)/2$ and contribute
	nonpositively to $\SC(\alternative_1)-\SC(\alternative_2)$.
	Voters in $\voters_{\alternative_3 \succ \alternative_2 \succ \alternative_1}$
	have distance to $\alternative_2$ at least
	$d(\alternative_3,\alternative_2)/2$.
	For a voter $i$ in
	$\voters_{\alternative_2 \succ \alternative_1 \succ \alternative_3}$,
	the triangle inequality and the ranking give
	\[
	\begin{aligned}
		d(i,\alternative_1)&\le d(i,\alternative_3)
		\le d(i,\alternative_2)+d(\alternative_3,\alternative_2),\\
		d(i,\alternative_2)&\ge
		\frac{d(\alternative_1,\alternative_2)-d(\alternative_3,\alternative_2)}{2}.
	\end{aligned}
	\]
	If $d(\alternative_3,\alternative_2)\ge
	d(\alternative_1,\alternative_2)$, then
	\[
	2\SC(\alternative_2)\ge
	(x_1+x_4)d(\alternative_1,\alternative_2),
	\]
	whereas
	\[
	\begin{aligned}
		\SC(\alternative_1)-\SC(\alternative_2)
		&\le(x_2+x_3+x_4)d(\alternative_1,\alternative_2)\\
		&\le\tau(x_1+x_4)d(\alternative_1,\alternative_2).
	\end{aligned}
	\]
	If $d(\alternative_3,\alternative_2)<
	d(\alternative_1,\alternative_2)$, then
	\[
	\begin{aligned}
		2\SC(\alternative_2)
		&\ge x_1d(\alternative_1,\alternative_2)
		+x_4d(\alternative_3,\alternative_2)\\
		&\quad+x_2\bigl(d(\alternative_1,\alternative_2)
		-d(\alternative_3,\alternative_2)\bigr)\\
		&=(x_1+x_2)\bigl(d(\alternative_1,\alternative_2)
		-d(\alternative_3,\alternative_2)\bigr)\\
		&\quad+(x_1+x_4)d(\alternative_3,\alternative_2),
	\end{aligned}
	\]
	and
	\[
	\begin{aligned}
		\SC(\alternative_1)-\SC(\alternative_2)
		&\le(x_3+x_4)d(\alternative_1,\alternative_2)
		+x_2d(\alternative_3,\alternative_2)\\
		&=(x_3+x_4)\bigl(d(\alternative_1,\alternative_2)
		-d(\alternative_3,\alternative_2)\bigr)\\
		&\quad+(x_2+x_3+x_4)d(\alternative_3,\alternative_2)\\
		&\le\tau\Bigl((x_1+x_2)\bigl(d(\alternative_1,\alternative_2)
		-d(\alternative_3,\alternative_2)\bigr)\\
		&\qquad\qquad+(x_1+x_4)d(\alternative_3,\alternative_2)\Bigr)\\
		&\le2\tau\SC(\alternative_2).
	\end{aligned}
	\]
	Both cases prove the claim without dividing by a social cost, so zero-cost
	optima are covered as well.
\end{proof}

We now describe the online rule. When the second alternative arrives, retain the alternative preferred by at least half of the voters, breaking ties deterministically. Denote the retained alternative by $\alternative_1$ and the discarded one by $\alternative_2$. When the third alternative $\alternative_3$ arrives, retain $\alternative_1$ if
\[
 |\voters_{\alternative_3 \succ \alternative_1}|\le\frac{\sqrt{13}-1}{2}\bigl(n-|\voters_{\alternative_3 \succ \alternative_1 \succ \alternative_2}|-|\voters_{\alternative_3 \succ \alternative_2 \succ \alternative_1}|\bigr),
\]
and otherwise select $\alternative_3$. Here $n-|\voters_{\alternative_3 \succ \alternative_1 \succ \alternative_2}|-|\voters_{\alternative_3 \succ \alternative_2 \succ \alternative_1}|$ is the number of voters who do not rank $\alternative_3$ first. Thus the final decision uses both the pairwise support of $\alternative_3$ against $\alternative_1$ and its first-place support, including its position relative to the discarded alternative $\alternative_2$.

\begin{observation}\label{pairwise-majority-bound}
	If $|\voters_{\alternative_1 \succ \alternative_2}|
	\geq |\voters_{\alternative_2 \succ \alternative_1}|$,
	then $SC(\alternative_1) \leq 3SC(\alternative_2)$.
\end{observation}

\begin{proof}
	By the triangle inequality, for every voter
	$v\in\voters_{\alternative_1 \succ \alternative_2}$,
	we have
	\[
	d(\alternative_1,\alternative_2)
	\leq d(v,\alternative_1)+d(v,\alternative_2)
	\leq 2d(v,\alternative_2).
	\]
	Therefore,
	\[
	\begin{aligned}
		SC(\alternative_2)
		&\geq |\voters_{\alternative_1 \succ \alternative_2}|
		\frac{d(\alternative_1,\alternative_2)}{2}\\
		&\geq |\voters_{\alternative_2 \succ \alternative_1}|
		\frac{d(\alternative_1,\alternative_2)}{2}.
	\end{aligned}
	\]
	Moreover, by the triangle inequality,
	\[
	\begin{aligned}
		SC(\alternative_1)-SC(\alternative_2)
		&\leq |\voters_{\alternative_2 \succ \alternative_1}|
		d(\alternative_1,\alternative_2)\\
		&\leq 2SC(\alternative_2).
	\end{aligned}
	\]
	Hence, $SC(\alternative_1)\leq 3SC(\alternative_2)$,
	completing the proof.
\end{proof}


\begin{lemma}[Three alternatives, metric, adversarial]\label{lem:three-uniform}
For every $n\ge1$, the above deterministic online rule has metric distortion at most $\sqrt{13}$ and runs in $O(n)$ time. Together with Lemma~\ref{lem:metric-det-lower}, this establishes $\sqrt{13}$ as the optimal uniform deterministic online distortion for three alternatives.
\end{lemma}
\begin{proof}
	Write $\tau=(\sqrt{13}-1)/2$, so $1<\tau<2$, $\tau^2+\tau=3$, and $1+2\tau=\sqrt{13}$. At the second arrival retain a pairwise-majority winner, breaking ties deterministically. Call the retained alternative $\alternative_1$ and the discarded one $\alternative_2$. When the third alternative $\alternative_3$ arrives, let
	\[
	q=|\voters_{\alternative_3 \succ \alternative_1}|,\qquad r=n-|\{\voter:\alternative_3\text{ is ranked first by }\voter\}|.
	\]
	Retain $\alternative_1$ if $q\le\tau r$, and otherwise select $\alternative_3$. The comparison can be implemented exactly using integers: switch if and only if $q^2+qr>3r^2$. The rule needs no distances and no information about future rankings.
	
	Let $x_1,\ldots,x_6$ be the numbers of voters with respective rankings
	\[
	\alternative_1 \succ \alternative_2 \succ \alternative_3,\quad \alternative_1 \succ \alternative_3 \succ \alternative_2,\quad \alternative_2 \succ \alternative_1 \succ \alternative_3,\quad \alternative_2 \succ \alternative_3 \succ \alternative_1,\quad \alternative_3 \succ \alternative_1 \succ \alternative_2,\quad \alternative_3 \succ \alternative_2 \succ \alternative_1.
	\]
	The second-step majority implies
	\[
	M=x_1+x_2+x_5-x_3-x_4-x_6\ge0.
	\]
	For use below, define
	\[
	\begin{aligned}
		F&=x_4+x_5+x_6-\tau(x_1+x_2+x_3+x_4),\\
		G&=x_1+x_3+x_4-\tau(x_1+x_2+x_5+x_6),\\
		H&=x_1+x_3-\tau(x_2+x_4+x_5+x_6).
	\end{aligned}
	\]
	The rule retains $\alternative_1$ exactly when $F\le0$.
	
	If it retains $\alternative_1$, its cost is at most $3\SC(\alternative_2)$ by the observation~\ref{pairwise-majority-bound}. To compare $\alternative_1$ with $\alternative_3$, apply Lemma~\ref{lem:three-comparison} with $(\alternative_1,\alternative_3,\alternative_2)$. Its first condition is $F\le0$. Its second condition follows from
	\[
	x_4+x_6\le x_1+x_2+x_5-x_3
	\le x_1+x_2+x_3+x_5
	\le\tau(x_1+x_2+x_3+x_5).
	\]
	Thus $\SC(\alternative_1)\le\sqrt{13}\SC(\alternative_3)$ as well.
	
	If the rule selects $\alternative_3$, then $F>0$ implies $x_4+x_5+x_6>x_1+x_2+x_3$. Thus a strict majority prefers $\alternative_3$ to $\alternative_1$, giving $\SC(\alternative_3)\le3\SC(\alternative_1)$ by observation~\ref{pairwise-majority-bound}. To compare $\alternative_3$ with $\alternative_2$, the two conditions of Lemma~\ref{lem:three-comparison}, with $(\alternative_3,\alternative_2,\alternative_1)$, are $G\le0$ and $H\le0$. The following exact identities prove them:
	\[
	\begin{aligned}
		F+H={}&-(\tau-1)(x_1+x_3+x_5+x_6)\\
		&-2\tau x_2-(2\tau-1)x_4\le0,
	\end{aligned}
	\]
	so $F>0$ gives $H<0$; and, using $\tau^2+\tau=3$,
	\[
	\begin{aligned}
		(\tau-1)F+(2-\tau)G+2(\tau-1)M
		={}&-(2-\tau)x_2-(\tau-1)x_3\\
		&-4(\tau-1)x_6\le0.
	\end{aligned}
	\]
	Since $M\ge0$, $F>0$, and both $\tau-1$ and $2-\tau$ are positive, the last identity gives $G<0$. Hence $\SC(\alternative_3)\le\sqrt{13}\SC(\alternative_2)$. This completes the proof.
\end{proof}

\subsection{Randomized Rules}
In this section, we investigate whether randomized voting rules admit improved distortion guarantees in the online setting. Using Lemma \ref{lem:gen:random}, we obtain a randomized algorithm with distortion $3-\frac{2}{n}$ based on Random Dictatorship. The rule chooses a voter uniformly at random and selects her top-ranked alternative.

\begin{lemma}\label{lem:random-dictatorship-regular}
	The Random Dictatorship rule is regular.
\end{lemma}

\begin{proof}
	Fix an alternative \(a \in \mathcal{A}\), and let \(\nu_a\) denote the number of voters who rank \(a\) first in the election \(\mathcal{E}\). Under Random Dictatorship, one voter is chosen uniformly at random and that voter's first-ranked alternative is selected. Therefore, the probability that \(a\) is selected is
	\[
	p_a=\frac{\nu_a}{n}.
	\]
	
	Now consider the extended election \(\mathcal{E}^+\), obtained by adding a new alternative \(a^+\). Let \(\nu'_a\) denote the number of voters who rank \(a\) first in \(\mathcal{E}^+\). Since introducing a new alternative can only cause some voters to move away from \(a\), we have
	\[
	\nu'_a \le \nu_a.
	\]
	The probability that \(a\) is selected in \(\mathcal{E}^+\) is
	\[
	p'_a=\frac{\nu'_a}{n}.
	\]
	Hence,
	\[
	p'_a=\frac{\nu'_a}{n}\le \frac{\nu_a}{n}=p_a.
	\]
	
	Thus, the selection probability of every existing alternative weakly decreases when the new alternative is added. Since this holds for every existing alternative \(a \in \mathcal{A}\), Random Dictatorship is regular.
\end{proof}

Now, combining Lemma \ref{lem:gen:random} with Lemma \ref{lem:random-dictatorship-regular} implies Corollary \ref{cor:olig}.
\begin{corollary}[of Lemma \ref{lem:random-dictatorship-regular}]
	\label{cor:olig}
	For the online voting problem under the metric assumption and adversarial arrival order of alternatives, there exists a randomized algorithm whose distortion is at most
	$
	3-\frac{2}{n}.
	$

\end{corollary}
\begin{proof}
For each voter $v$, let $b_v$ be her favorite alternative and let $a^*$ minimize social cost. For every other voter $u$, the triangle inequality and $d(v,b_v)\le d(v,a^*)$ imply $d(u,b_v)\le d(u,a^*)+2d(v,a^*)$. For $u=v$, use $d(v,b_v)\le d(v,a^*)$. Hence $\SC(b_v)\le\SC(a^*)+2(n-1)d(v,a^*)$. Averaging uniformly over $v$ proves the  factor $3-2/n$.
\end{proof}

Finally, it is natural to ask whether one can improve upon the distortion bound of $3$ and achieve a guarantee of the form $3-\varepsilon$ for some constant $\varepsilon>0$. In the offline setting, Charikar et al. \cite{charikar2024breaking} recently broke the long-standing barrier of $3$ by proposing a randomized voting rule with distortion $2.753$. This raises the question of whether similar improvements can be achieved in the online setting.

Existing approaches that obtain distortion strictly below $3$ typically combine several voting rules. A key ingredient in these constructions is the Maximal Lotteries rule. Thus, it is natural to investigate whether Maximal Lotteries can be implemented online while preserving regularity.

\begin{definition}The Maximal Lotteries (ML) rule is defined through a zero-sum game between two players, Alice and Bob, over the set of candidates. In this game, both players simultaneously choose a distribution over alternatives. Let \(D_A\) and \(D_B\) denote their respective mixed strategies. A pair of candidates \(a\) and \(b\) is then sampled independently according to \(D_A\) and \(D_B\). Alice receives payoff \(1\) if \(a\) beats \(b\) in a pairwise majority comparison, Bob receives payoff \(1\) if \(b\) beats \(a\), and both players receive payoff \(1/2\) if \(a=b\).
	Maximal Lottery returns the distribution \(D^*\) that arises as a Nash equilibrium of this game. The ML rule then outputs a single alternative by sampling according to \(D^*\). 
\end{definition}

Charikar et al. \cite{charikar2024breaking} show that we can combine this rule with another rule and obtain a distortion which is strictly less than $3$.

Here, we show that the ML rule is not regular. 

\begin{table}[t]
	\centering
	\begin{subtable}[t]{0.45\textwidth}
		\centering
		\begin{tabular}[t]{c|c|c|c}
			& $\voter_1$ & $\voter_2$ & $\voter_3$ \\ \hline
			\text{1st} & $\alternative_1$  & $\alternative_1$ & $\alternative_2$ \\
			\text{2nd} & $\alternative_2$  & $\alternative_2$ & $\alternative_1$ 
		\end{tabular}
		\caption{}
		\label{table:ml-counterexample-a}
	\end{subtable}
	\hspace{1em}          
	\begin{subtable}[t]{0.45\textwidth}
		\centering
		\begin{tabular}[t]{c|c|c|c}
			& $\voter_1$ & $\voter_2$ & $\voter_3$ \\ \hline
			\text{1st} & $\alternative_1$  & $\alternative_3$ & $\alternative_2$ \\
			\text{2nd} & $\alternative_2$  & $\alternative_1$ & $\alternative_3$ \\
			\text{3rd} & $\alternative_3$  & $\alternative_2$ & $\alternative_1$
		\end{tabular}
		\caption{}
		\label{table:ml-counterexample-b}
	\end{subtable}
	\caption{Nash equilibrium in Table~\ref{table:ml-counterexample-a} is $(1 , 0)$ and in Table~\ref{table:ml-counterexample-b} is $(\frac{1}{3},\frac{1}{3},\frac{1}{3})$}
	\label{table:ml-counterexample}
\end{table}

\begin{lemma}\label{lem:maxLotteryNotMonotone}
	The Maximal Lottery rule is not regular.
\end{lemma}

\begin{proof}
	We show that the Maximal Lottery rule is not regular by constructing a counter example.
	Consider the profile in Table~\ref{table:ml-counterexample-a}, which contains only two alternatives, \(\alternative_1\) and \(\alternative_2\). In this profile, two out of three voters prefer \(\alternative_1\) to \(\alternative_2\), so \(\alternative_1\) is the Condorcet winner, and the Maximal Lottery rule always selects the Condorcet winner with probability \(1\). Hence, in this profile, \(\alternative_2\) is selected with probability \(0\).
	Next, we extend this profile by introducing a third alternative \(\alternative_3\), as shown in Table~\ref{table:ml-counterexample-b}. The addition of \(\alternative_3\) creates a rock-paper-scissors structure: a majority prefers \(\alternative_1\) to \(\alternative_2\), \(\alternative_2\) to \(\alternative_3\), and \(\alternative_3\) to \(\alternative_1\).  In this case, the Nash equilibrium is the uniform distribution over the three alternatives. Therefore, the Maximal Lottery rule selects each alternative, including \(\alternative_2\), with probability \(1/3\).
	
	Comparing the two cases, we see that the probability of selecting \(\alternative_2\) increases from \(0\) to \(1/3\) after adding the new alternative \(\alternative_3\). Hence, the Maximal Lottery rule is not regular.
\end{proof}

	\section{Conclusion}
In this paper, we initiated the study of distortion in online voting, investigating both deterministic and randomized mechanisms under utilitarian and metric objectives. We established upper and lower bounds across these settings and introduced a regularity-based framework for converting offline randomized voting rules into online ones while preserving their outcome distributions.

Our results leave several interesting questions open. First, in the randomized utilitarian setting, there remains a gap between the known lower bound of \(\Omega(\sqrt{m})\) and our upper bound of \(O(\sqrt{m\log m})\). Can this gap be closed, eliminating the logarithmic factor and achieving an \(O(\sqrt{m})\) distortion guarantee in the online setting?

Second, in the deterministic metric setting, we establish an upper bound of \(2+\sqrt{5}\) for an arbitrary number of alternatives, while For three alternatives, we established a tight distortion bound of \(\sqrt{13}\), over all electorate sizes. An interesting open question is whether this guarantee can be extended to elections with more than three alternatives, or whether additional alternatives necessarily lead to higher distortion.

Finally, in the randomized metric setting, our Random Dictatorship-based mechanism achieves a distortion of at most \(3-\frac{2}{n}\), whereas the known lower bound is \(2.1126\). Can we design an online randomized mechanism that achieves a constant distortion bounded away from \(3\), and potentially closes this gap?

	\FloatBarrier
	
	\bibliographystyle{unsrt}
	\bibliography{draft}

\end{document}